\documentclass[10pt,aps,showpacs,nofootinbib,prd,aps,epsf,floats,
               amsmath,amssymb,amsfonts,axodraw,twocolumn]{revtex4}
\usepackage{amsmath, amssymb}
\newcommand{\mathsym}[1]{{}}

\usepackage{graphicx}
\usepackage{amsmath}
\usepackage{amssymb}
\usepackage{bm}
\newcommand{\be}{\begin{equation}}
\newcommand{\ee}{\end{equation}}
\newcommand{\bea}{\begin{eqnarray}}
\newcommand{\eea}{\end{eqnarray}}

\newcommand{\rem}[1]{}
\newsavebox{\PSLASH}
 \sbox{\PSLASH}{$p$\hspace{-1.8mm}/}
 
\renewcommand{\theequation}{\thesection.\arabic{equation}}
\newcounter{saveeqn}
\newcommand{\add}{\addtocounter{equation}{1}}
\newcommand{\alpheqn}{\setcounter{saveeqn}{\value{equation}}%
\setcounter{equation}{0}%
\renewcommand{\theequation}{\mbox{\thesection.\arabic{saveeqn}{\alph{equation}}}}}
\newcommand{\reseteqn}{\setcounter{equation}{\value{saveeqn}}%
\renewcommand{\theequation}{\thesection.\arabic{equation}}}

 \newsavebox{\notrightarrow}
 \sbox{\notrightarrow}{$\to$\hspace{-4mm}/}
 
 \newsavebox{\PARTIALSLASH}
 \sbox{\PARTIALSLASH}{$\partial$\hspace{-1.6mm}/}
 
 \newsavebox{\ASLASH}
 \sbox{\ASLASH}{$A$\hspace{-2.1mm}/}
 
 \newsavebox{\KSLASH}
 \sbox{\KSLASH}{$k$\hspace{-1.8mm}/}
 
 \newsavebox{\LSLASH}
 \sbox{\LSLASH}{$\ell$\hspace{-1.8mm}/}
 
 \newsavebox{\QSLASH}
 \sbox{\QSLASH}{$q$\hspace{-1.8mm}/}
 
 \newsavebox{\DSLASH}
 \sbox{\DSLASH}{$D$\hspace{-2.2mm}/}
 
 \newsavebox{\DbfSLASH}
 \sbox{\DbfSLASH}{${\mathbf D}$\hspace{-2.8mm}/}
 
 \newsavebox{\DELVECRIGHT}
 \sbox{\DELVECRIGHT}{$\stackrel{\rightarrow}{\partial}$}
 
 \newcommand{\blue}{\IfColor{\textCadetBlue}{}}
\newcommand{\black}{\IfColor{\textBlack}{}}
\newcommand{\red}{\IfColor{\textRed}{}}
\newcommand{\green}{\IfColor{\textOliveGreen}{}}
\newcommand{\lila}{\IfColor{\textRedViolet}{}}

\begin{document}
\title{Local electric current correlation function in\\ an exponentially decaying magnetic field}
\author{N. Sadooghi}\email{sadooghi@physics.sharif.ir}
\author{F. Taghinavaz}\email{taghinavaz@physics.sharif.ir}
\affiliation{Department of Physics, Sharif University of Technology,
P.O. Box 11155-9161, Tehran-Iran}
\begin{abstract}
\noindent The effect of an exponentially decaying magnetic field on
the dynamics of Dirac fermions in $3+1$ dimensions is explored. The
spatially decaying magnetic field is assumed to be aligned in the
third direction, and is defined by
${\mathbf{B}}(x)=B(x)\hspace{0.05cm}{\mathbf{e}}_{z}$, with
$B(x)=B_{0}e^{-\xi x/\ell_{B}}$. Here, $\xi$ is a dimensionless
damping factor and $\ell_{B}=(eB_{0})^{-1/2}$ is the magnetic
length. As it turns out, the energy spectrum of fermions in this
inhomogeneous magnetic field can be analytically determined using
the Ritus method. Assuming the magnetic field to be strong, the
chiral condensate and the \textit{local} electric current
correlation function are computed in the lowest Landau level (LLL)
approximation and the results are compared with those arising from a
strong homogeneous magnetic field. Although the constant magnetic
field $B_{0}$ can be reproduced by taking the limit of $\xi\to 0$
and/or $x\to 0$ from $B(x)$, these limits turn out to be singular,
especially once the quantum corrections are taken into account.
\end{abstract}
 \pacs{11.15.-q, 11.30.Qc, 11.30.Rd, 12.20.-m, 12.20.Ds} \maketitle
\section{Introduction}\label{introduction}
\noindent The effects of magnetic fields on systems containing
relativistic fermions are subject of intense theoretical studies
over the past two decades. These effects include a wide range, from
condensed matter physics \cite{condensed} to high energy physics
\cite{hep} and cosmology \cite{cosmology}. Theoretical studies deal
in the most of these works with the idealized limit of constant and
homogeneous magnetic fields. However, this limit is only reliable as
long as the scale of field variation is much larger than the Compton
wavelength of the fermionic system. As it is shown in
\cite{klimenko1992}, strong and homogeneous magnetic fields lead to
the formation of fermion bound states even in the weakest attractive
interaction between the fermions. This phenomenon, known as magnetic
catalysis of dynamical chiral symmetry breaking \cite{miransky1995},
is essentially based on a dimensional reduction from $D$ to $D-2$
dimensions in the presence of strong homogeneous magnetic fields in
the regime of LLL dominance. However, as it is demonstrated recently
in \cite{martino2007}, massless Dirac electrons can also be confined
by inhomogeneous magnetic fields in graphene, which is a single
layer of carbon atoms in a honeycomb lattice. Electrons in graphene
are described by a massless two-dimensional, time-independent
relativistic Dirac equation. As it is shown in \cite{martino2007},
magnetic quantum dots can be formed, if one neglects the effects of
electron spin and solves the Dirac equation in the presence of a
square well magnetic barrier of width $2d$,
${\mathbf{B}}(\mathbf{x})=B(x)\hspace{0.05cm}{\mathbf{e}}_{z}$,
directed perpendicular to the graphene $x-y$ plane, with
$B(x)=B_{0}\theta(d^{2}-x^{2})$, and $\theta$ the Heaviside step
function. A circularly symmetric magnetic quantum dot arises then by
a radially inhomogeneous magnetic field
${\mathbf{B}}=B(r)\hspace{0.03cm}{\mathbf{e}}_{z}$ (see
\cite{martino2007, kormanyos2008} for more details). In
\cite{roy2011}, the bound state solutions and the spectra of
graphene excitations in the presence of an inverse radial magnetic
field, with $B(r)\sim\frac{1}{r}$, is studied. Analytical solutions
for the Dirac equation in the presence of double or multiple
magnetic barriers are presented in \cite{doublebarrier}. The bound
spectra of electrons in a magnetic barrier with hyperbolic profile
is studied in \cite{murguia2011}.
\par
In the present paper, we will focus, in particular, on the solution
of Dirac equation in the presence of an exponentially decaying
magnetic field. Analytical solution of a quasi-two-dimensional Dirac
equation including the spin of a single electron in a magnetic field
of constant direction with arbitrarily strong exponentially
depending variation perpendicular to the field direction is firstly
presented in \cite{handrich2005}. In \cite{ghosh2009}, the
probability density and current distributions of Dirac electrons in
graphene in the presence of an exponentially decaying magnetic field
are determined. The formation of chiral condensate induced by
exponentially decaying magnetic fields, in $2+1$ dimensions, is
studied in \cite{raya2010}. The vacuum polarization tensor of $3+1$
dimensional scalar Quantum Electrodynamics in the presence of an
inhomogeneous background magnetic field, arising from a general
plane wave basis for the underlying gauge field, is computed
recently in \cite{gies2011}. In order to demonstrate the
\textit{nonlocal} features of quantum field theory in the presence
of inhomogeneous background magnetic fields, the vacuum polarization
tensor is defined to be \textit{local}. This turns out to be a
useful task providing \textit{local} information about the
\textit{nonlocal} nature of fluctuation-induced processes
\cite{gies2011}. In the present paper, following the same idea as in
\cite{gies2011}, we will define a \textit{local} electric current
correlation function (local electric susceptibility), and compute it
in the presence of a strong and exponentially decaying magnetic
field in the LLL approximation. Eventually, we will compare the
result with the electric current correlation function arising from a
strong homogeneous magnetic field and discuss the effects of
inhomogeneity of the background magnetic field.
\par
The organization of the paper is as follows: In Sec. \ref{sec2a}, we
will briefly review the Ritus eigenfunction method \cite{ritus} by
solving the Dirac equation in the presence of a constant and
homogeneous magnetic field. In Sec. \ref{sec2b}, we will introduce
an exponentially decaying magnetic field
${\mathbf{B}}(x)=B(x)\hspace{0.03cm}{\mathbf{e}}_{z}$, which is
assumed to be aligned in the third direction, with $B(x)$ given by
$B(x)=B_{0}e^{-\xi x/\ell_{B}}$. Here, $\xi$ is a dimensionless
damping parameter and $\ell_{B}=(eB_{0})^{-1/2}$ is the magnetic
length. Using the Ritus method, the energy eigenvalues and
eigenfunctions of a $3+1$ dimensional Dirac equation in the presence
of ${\mathbf{B}}(x)$ will be determined. In Sec. \ref{sec3}, we will
determine the chiral condensate $\langle\bar{\psi}\psi\rangle$,
using the fermion propagator in the presence of strong homogeneous
(Sec. \ref{sec3a}) and exponentially decaying magnetic fields (Sec.
\ref{sec3b}) in the LLL approximation. We will show that whereas
$\langle\bar{\psi}\psi\rangle$ in a strong homogeneous magnetic
field is constant, $\langle\bar{\psi}\psi\rangle$ in an
exponentially decaying magnetic field depends nontrivially on a
dimensionless variable $u$, defined by $u=\frac{2}{\xi^{2}}e^{-\xi
x/\ell_{B}}$. In Sec. \ref{sec4}, the local electric current
correlation function (local electric susceptibility) $\chi^{(i)}$
will be computed in the presence of strong homogeneous (Sec.
\ref{sec4a}) and exponentially decaying magnetic fields (Sec.
\ref{sec4b}) in the LLL approximation. Here, $i=1,2,3$ denotes the
three spatial directions. As it turns out, in the regime of LLL
dominance the electric susceptibility in the perpendicular direction
relative to the external magnetic field (i.e. for $i=1,2$) vanishes
in both homogeneous and inhomogeneous magnetic fields. The
longitudinal component of $\chi^{(i)}$, denoted by $\chi^{\|}$,
turns out to be constant in a homogeneous magnetic field and depend
nontrivially on the dimensionless variable $u$ in an exponentially
decaying magnetic field. In Sec. \ref{sec5}, we will compare the
results for the chiral condensate and the local electric current
correlation function arising from homogeneous and inhomogeneous
magnetic fields and show that the limits $\xi\to 0$ as well as $x\to
0$ are singular. In other words, although in the limits $\xi\to 0$
and/or $x\to 0$, we get $B(x)\to B_{0}$, but the values of
$\langle\bar{\psi}\psi\rangle$ and $\chi^{\|}$ arising from an
exponentially decaying magnetic field  in these limits are not the
same as $\langle\bar{\psi}\psi\rangle$ and $\chi^{\|}$ arising from
a homogeneous magnetic field. This demonstrates the nonlocal
features of the quantum vacuum in a strong and spatially decaying
magnetic field. As is indicated in Sec. \ref{sec6}, which is devoted
to our concluding remarks, the results of this paper are relevant
for high energy physics as well as condensed matter physics.
\section{Electrons in external magnetic fields}\label{sec2}
\setcounter{equation}{0}\par\noindent Ritus eigenfunction method
\cite{ritus} is a powerful tool to study the dynamics of electrons
in external electromagnetic fields. It is recently used in
\cite{murguia2009} to solve the Dirac equation in the presence of a
uniform magnetic field in 2+1 dimensions, a uniform electric field
in 1+1 dimensions and an exponentially decaying magnetic field in
$2+1$ dimensions (see also \cite{raya2010}). In \cite{ritus-four,
fukushima2009}, the Ritus method is used to solve the $3+1$
dimensional Dirac equation for massless fermions in the presence of
a constant magnetic field. In Sec. \ref{sec2a}, we will briefly
review the results from \cite{fukushima2009} and will present the
solution of the Dirac equation for a single massive electron in the
presence of a constant magnetic field in $3+1$ dimensions. This will
fix our notations. In Sec. \ref{sec2b}, we will use the same method
and will solve the Dirac equation for a massive electron in an
exponentially decaying magnetic field. Our results coincide with the
result presented in \cite{raya2010}, although our notations are
slightly different. We will also present the final form of the
fermion propagator in the presence of a uniform and an exponentially
decaying magnetic field in Secs. \ref{sec2a} and \ref{sec2b},
respectively. The results presented in this section will then be
used in Sec. \ref{sec3} to determine the chiral condensate, and in
Sec. \ref{sec4} the local electric current correlation function in
the presence of uniform and non-uniform magnetic fields.
\subsection{Electrons in
constant magnetic fields}\label{sec2a}
\setcounter{equation}{0}\par\noindent Let us start with the Dirac
equation of a single electron in $3+1$ dimensional Minkowski space
in a background electromagnetic field
\begin{eqnarray}\label{A1}
\left(\gamma\cdot \Pi-m\right)\psi=0,
\end{eqnarray}
where $\Pi_{\mu}\equiv i\partial_{\mu}-eA_{\mu}$ and $A_{\mu}$ is
the electromagnetic potential. Here, $m$ and $e$ are the electron's
mass and electric charge, respectively. The aim is to solve
(\ref{A1}) and to eventually determine the fermion propagator in the
presence of background constant magnetic field
$\mathbf{B}=B_{0}\mathbf{e}_{z}$, aligned in the third direction.
Such a constant magnetic field, is built, for instance, by choosing
the gauge field $A_{\mu}$ in the Landau gauge as
$A_{\mu}=\left(0,0,B_{0}x,0\right)$.  To solve (\ref{A1}), we use,
as in \cite{ritus-four, murguia2009}, the Ansatz
$\psi=\mathbb{E}_{p}u_{\tilde{p}}$. Here, $\mathbb{E}_{p}$ is a
diagonal matrix satisfying
\begin{eqnarray}\label{A2}
(\gamma\cdot\Pi)~\mathbb{E}_{p}=\mathbb{E}_{p}~(\gamma\cdot\tilde{p}),
\end{eqnarray}
and $u_{\tilde{p}}$ is a free spinor, that describes an electron
with momentum $\tilde{p}$ and satisfies
$(\gamma\cdot\tilde{p}-m)u_{\tilde{p}}=0$. The Ritus eigenfunction
$\mathbb{E}_{p}$ and the Ritus momentum $\tilde{p}$ in (\ref{A2})
are unknown and shall be determined in what follows. The matrix
$\mathbb{E}_{p}$ is determined by solving the eigenvalue equation
\begin{eqnarray}\label{A3}
(\gamma\cdot\Pi)^{2}~\mathbb{E}_{p}=\tilde{p}^{2}\mathbb{E}_{p},
\end{eqnarray}
that arises directly from (\ref{A2}). Using
$(\gamma\cdot\Pi)^{2}=\Pi^{2}+\frac{e}{2}\sigma^{\mu\nu}F_{\mu\nu}$,
with $\sigma^{\mu\nu}\equiv \frac{i}{2}[\gamma^{\mu},\gamma^{\nu}]$
and the field strength tensor $F_{\mu\nu}\equiv
\partial_{\mu}A_{\nu}-\partial_{\nu}A_{\mu}$, with non-vanishing elements $F_{12}=-F_{21}=B_{0}$,
and choosing, without loss of generality, the Dirac
$\gamma$-matrices as $\gamma^{0}=\sigma_{3}\otimes \sigma_{3}$, and
$\gamma^{i}=i\sigma_{3}\otimes \sigma_{i}$, where $\sigma_{i},
i=1,2,3$ are the Pauli matrices, (\ref{A3}) reads
\begin{eqnarray}\label{A4}
\left(\Pi^{2}+\sigma^{12}
eB_{0}\right){\mathbb{E}}_{p}=\tilde{p}^{2}\mathbb{E}_{p}.
\end{eqnarray}
Here, $\sigma^{12}=i\gamma^{1}\gamma^{2}=\mathbb{I}\otimes
\sigma_{3}$ with $\mathbb{I}$ a $2\times 2$ identity matrix. To
solve (\ref{A4}), we use the fact that in the Landau gauge, the
operator $\left(\gamma\cdot\Pi\right)$ commutes with $\Pi_0$,
$\Pi_{j}, j=2,3$ and ${\cal{P}}\equiv
-\left(\gamma\cdot\Pi\right)^{2}+\Pi_{0}^{2}-\Pi_{3}^{2}$, and
therefore has simultaneous eigenfunctions with these operators. The
eigenvalues of these operators, defined by the eigenvalue equations
\begin{eqnarray}\label{A5}
\Pi_0\mathbb{E}_{p}&=&p_{0}\mathbb{E}_{p},\qquad\Pi_j\mathbb{E}_{p}=p_{j}\mathbb{E}_{p},~
j=2,3,\hspace{0.5cm}\mbox{and}\nonumber\\
{\cal{P}}\mathbb{E}_{p}&=&p\ \mathbb{E}_{p},
\end{eqnarray}
label the solutions of Dirac equation (\ref{A1}) in the background
magnetic field. Using (\ref{A3}), the definition of ${\cal{P}}$ and
(\ref{A5}), it can be shown that the four-momentum $\tilde{p}$
satisfies $\tilde{p}^{2}=p_{0}^{2}-p_{3}^{2}-p$, and
${\cal{P}}\mathbb{E}_{p}=p~\mathbb{E}_{p}$ is therefore given by
\begin{eqnarray}\label{A6}
\left(\partial_{x}^{2}-(p_{2}-eB_{0}x)^{2}+\sigma^{12}eB_{0}\right)\mathbb{E}_{p}=-p~\mathbb{E}_{p}.
\end{eqnarray}
Before solving (\ref{A6}), let us determine the components of Ritus
momentum $\tilde{p}$, which turns out to play an important role in
this method. To do this, we consider (\ref{A2}), and use the Ansatz
\begin{eqnarray}\label{A7}
\mathbb{E}_{p}=\sum\limits_{\sigma=\pm}\Delta_{\sigma}E_{p,\sigma},
\end{eqnarray}
with the projectors \cite{ritus-four}
\begin{eqnarray}\label{A8}
\Delta_{\sigma}&\equiv&\mbox{diag}\left(\delta_{\sigma,1},\delta_{\sigma,-1},\delta_{\sigma,1},\delta_{\sigma,-1}\right).
\end{eqnarray}
Plugging first $\mathbb{E}_{p}$ from (\ref{A7}) in the left hand
side (l.h.s) of (\ref{A2}), and then comparing the resulting
expression with the expression arising from the right hand side
(r.h.s.) of (\ref{A2}), where again $\mathbb{E}_{p}$ from (\ref{A7})
and the Ansatz $\tilde{p}=(p_{0},0,\tilde{p}_{2},p_{3})$ is
used,\footnote{The form $\tilde{p}^{2}=p_{0}^{2}-p_{3}^{2}-p$
suggests that $\tilde{p}_{1}=0$. Moreover, we have fixed
$\tilde{p}_{i}=p_{i}, i=0,3$, where according to (\ref{A5}) $p_{i},
i=0,3$ are the eigenvalue of $\Pi_{i}, i=0,3$, respectively.} we
arrive at (\ref{A6}), with $p$ given by $p=\tilde{p}_{2}^{2}$. This
will fix $\tilde{p}_{2}=-\sqrt{p}$.\footnote{Note that the general
solution for $\tilde{p}_{2}$ is either $+\sqrt{p}$ or $-\sqrt{p}$.
Here, we have chosen $\tilde{p}_{2}=-\sqrt{p}$ to have the same
notation as in \cite{fukushima2009}.}
\par
Let us now turn back to the solution of (\ref{A6}), whose form
suggests a plane wave Ansatz in the $0,2$ and $3$ directions,
\begin{eqnarray}\label{A12}
E_{p,\pm 1}(\bar{x})=e^{-i\bar{p}\cdot \bar{x}}f^{\pm}_{p}(x),
\end{eqnarray}
where $\bar{x}=(t,x,y,z)$ and $\bar{p}=(p_{0},0,p_{2},p_{3})$.
Plugging (\ref{A12}) in (\ref{A7}), and using
$\Delta_{\pm}=\frac{1}{2}\left(1\pm i\gamma^{1}\gamma^{2}\right)$
for the projectors defined in (\ref{A8}), the general solution for
$\mathbb{E}_{p}$ is given by
\begin{eqnarray}\label{A13}
\mathbb{E}_{p}(\bar{x})=e^{-i(p_{0}t-p_{2}y-p_{3}z)}P_{p}(x),
\end{eqnarray}
where
\begin{eqnarray}\label{A14}
\lefteqn{\hspace{-1cm}P_{p}(x)=}\nonumber\\
&&\hspace{-1.2cm}\frac{1}{2}\{[f_{p}^{+}(x)+f_{p}^{-}(x)]+
i\gamma^{1}\gamma^{2}[f_{p}^{+}(x)-f_{p}^{-}(x)]\}.
\end{eqnarray}
To determine $f_{p}^{\pm}(x)$, it is enough to replace
$\mathbb{E}_{p}(\bar{x})$ from (\ref{A13}) in (\ref{A6}) to arrive
at
\begin{eqnarray}\label{A15}
\hspace{-0.5cm}\left(\partial_{x}^{2}-\left(p_{2}-eB_{0}x\right)^{2}\pm
eB_{0}\right)f_{p}^{\pm}(x)=-pf_{p}^{\pm}(x).
\end{eqnarray}
Renaming the discrete quantum numbers $\sqrt{p}$ in the
four-momentum $\tilde{p}$ by $\sqrt{p}=
\mbox{sgn}(eB_{0})\sqrt{2|eB_{0}|p}$, and choosing a new coordinate
$\zeta\equiv \frac{\sqrt{2}}{\ell_{B}}(x-\ell_{B}^{2}p_{2})$, with
$\ell_{B}$ the magnetic length defined by $\ell_{B}\equiv
|eB_{0}|^{-1/2}$, the differential equation in (\ref{A15}) can be
reformulated in the form of the differential equation of harmonic
oscillator, whose center lies in $\bar{x}_{0}=p_{2}/eB_{0}$ and
oscillates with the cyclotron frequency $\omega_{c}=2eB_{0}$
\cite{murguia2009},
\begin{eqnarray}\label{A16}
\left(\frac{\partial^{2}}{\partial\zeta^{2}}+p+\frac{\sigma}{2}\mbox{sgn}(eB_{0})-\frac{\zeta^{2}}{4}\right)f_{p}^{\pm}(\zeta)=0.
\end{eqnarray}
As it turns out, the solution to the above equation can be given in
terms of parabolic cylinder function \cite{ritus-four,murguia2009}
\begin{eqnarray}\label{A18}
\phi_{n_{\sigma}}(x)&=&a_{n_{\sigma}}\exp\left(-\frac{x^{2}}{2\ell_{B}^{2}}\right)H_{n_{\sigma}}\left(\frac{x}{\ell_{B}}\right),
\nonumber\\
\mbox{with}&&
a_{n_{\sigma}}\equiv\frac{1}{\sqrt{2^{n_{\sigma}}n_{\sigma}!\ell_{B}\sqrt{\pi}}},
\end{eqnarray}
where $n_{\sigma}$ is, in general, given by
\begin{eqnarray}\label{A18a}
n_{\sigma}=p+\frac{\sigma}{2}\mbox{sgn}(eB_{0})-\frac{1}{2}.
\end{eqnarray}
Here, $p$ labels the energy (Landau) levels. In the rest of this
paper, we will work with $eB_{0}>0$. According to (\ref{A18a}), for
sgn$(eB_{0})>0$, the condition that $n_{\sigma}\geq 0$ for $p=0$,
fixes the spin orientation of the electrons in the LLL to be
positive ($\sigma=+1$). In this case $n_{+}$ will be given by
$n_{+}=p$ for all values of $p$ \cite{ritus-four}. For this specific
choice, the final result for $f_{p}^{\pm}(x)$ is the same as the
result presented in \cite{fukushima2009},
\begin{eqnarray}\label{A17}
f_{p}^{+}(x)&=&\phi_{p}\left(x-\ell_{B}^{2} p_{2}\right),
\qquad\qquad
p=0,1,2,\cdots,\nonumber\\
f_{p}^{-}(x)&=&\phi_{p-1}\left(x-\ell_{B}^{2} p_{2}\right),
\qquad\hspace{0.35cm} p=1,2,3,\cdots.\nonumber\\
\end{eqnarray}
Hence, for $eB_{0}>0$, only the positive spin solution
$f_{0}^{+}(x)$ contributes in the LLL, characterized by $p=0$. In
other words, $f_{0}^{-}(x)=f_{-1}^{+}(x)$ is undefined and is to be
neglected, whenever it appears in a computation.
\par
The energy dispersion relation for an electron in the presence of a
constant magnetic field is determined by the Ritus momentum
\begin{eqnarray}\label{A20}
\tilde{p}_{p}=\left(p_{0},0,-\sqrt{2eB_{0}p},p_{3}\right),
\end{eqnarray}
and is given by
\begin{eqnarray}\label{A21}
\hspace{-0.3cm}{\cal{E}}_{p}=\sqrt{2eB_{0}p+p_{3}^{2}+m^{2}},\hspace{0.3cm}
p=0,1,2,\cdots.
\end{eqnarray}
The Ritus eigenfunction $\mathbb{E}_{p}$ from (\ref{A13}) can be
used to determine the fermion propagator in the presence of a
uniform magnetic field. To do this, the Ansatz
$\psi=\mathbb{E}_{p}u_{\bar{p}}$ for the electron is to be
generalized to a system including particles and antiparticles.
Defining the associated creation and annihilation operators, and
following the standard steps to determine the propagator
\cite{itzykson}, the fermion propagator in the presence of a
constant magnetic field is given by (see also \cite{fukushima2009})
\begin{eqnarray}\label{A22}
\lefteqn{G(\bar{x},\bar{x}')\equiv\langle\psi(\bar{x})\bar{\psi}(\bar{x}')\rangle}\nonumber\\
&&=\sum_{p=0}^{\infty}\int
{\cal{D}}\bar{p}~e^{-i\bar{p}\cdot(\bar{x}-\bar{x}')}
P_{p}(x)\frac{i}{(\gamma\cdot\tilde{p}_{p}-m)}~P_{p}(x'),\nonumber\\
\end{eqnarray}
where ${\cal{D}}\bar{p}\equiv
\frac{dp_{0}dp_{2}dp_{3}}{(2\pi)^{3}}$, $P_{p}(x)$ is given in
(\ref{A14}) with $f_{p}^{\pm}(x)$ from (\ref{A17}), and the Ritus
momentum  $\tilde{p}_{p}$ is given in (\ref{A21}). The same
expression appears also in \cite{fukushima2009}. Since we are
working with $eB_{0}>0$, and the spin orientation of the electrons
are already fixed in the LLL to be positive, no spin projector
appears in (\ref{A22}), as in the fermion propagators presented in
\cite{ritus-four}.  To understand this, let us notice again that the
facts that LLL includes only the positive spin solution, and that
negative spin solutions contribute only to the higher Landau levels,
are explicitly implemented in the choice $p=0,1,2,\cdots$ for
positive spin solution $f_{p}^{+}(x)$, and $p=1,2,3,\cdots$ for
negative spin solution $f_{p}^{-}(x)=f_{p-1}^{+}(x)$ in (\ref{A17}).
In Sec. \ref{sec2b}, we will use the above method to determine the
energy dispersion relation and the propagator for electrons in the
presence of exponentially decaying magnetic fields.
\subsection{Electrons in
exponentially decaying magnetic fields}\label{sec2b}
\par\noindent
In this section, the Ritus eigenfunction method will be used to
determine the energy levels of a single fermion in the presence of
an exponentially decaying magnetic field
${\mathbf{B}}(x)=B_{0}e^{-\alpha x}\hat{\mathbf{e}}_{z}$, aligned in
the third direction. Here, $B_{0}$ is the magnetic field at $x=0$,
$\alpha$ is a dimensionful damping parameter. Later, we will set
$\alpha=\xi\sqrt{eB_{0}}$, where $\xi$ is a given dimensionless
damping parameter and $eB_{0}>0$. The dynamics of this electron in
such an inhomogeneous magnetic field is described by the Dirac
equation (\ref{A1}). To solve it, we fix the gauge, in contrast to
the previous case, by
$A_{\mu}(x)=\left(0,0,-\frac{B_{0}}{\alpha}(e^{-\alpha
x}-1),0\right)$, as in \cite{raya2010}. With this choice of
$A_{\mu}$, the case of constant magnetic field will be recovered by
taking the limit $\alpha\to 0$ (or equivalently $\xi\to 0$) in the
classical level. Using the Ansatz
$\psi=\mathbb{E}_{\bar{p}}u_{\tilde{p}}$, with $\mathbb{E}_{p}$
satisfying (\ref{A2}), and following the same steps leading from
(\ref{A2}) to (\ref{A6}), we arrive at
\begin{eqnarray}\label{A23}
&&\hspace{-0.5cm}\bigg[\partial_{x}^{2}-\left(i\partial_{y}-\frac{eB_{0}}{\alpha}(e^{-\alpha
x}-1)\right)^{2}+\sigma eB_{0}e^{-\alpha x}\bigg]\mathbb{E}_{p}
\nonumber\\
&&\hspace{0.5cm}=-p~\mathbb{E}_{p},
\end{eqnarray}
where $\sigma=\pm 1$ are the eigenvalues of third Pauli matrix,
$\sigma_{3}=\mbox{diag}(1,-1)$. Using in analogy to (\ref{A7}), a
plane wave Ansatz for $\mathbb{E}_{p}$,\footnote{Note that the plane
wave Ansatz is justified by
$[{\cal{P}},i\partial_{3}]=[{\cal{P}},i\partial_{0}]=0$, where
${\cal{P}}$ is defined by
${\cal{P}}=-(\gamma\cdot\Pi)^{2}+\Pi_{0}^{2}-\Pi_{3}^{2}$, as in
Sec. \ref{sec2a}.}
\begin{eqnarray}\label{A24}
\mathbb{E}_{p}(\rho)=\sum\limits_{\sigma=\pm}\Delta_{\sigma}E_{p,\sigma}(\rho),
\end{eqnarray}
with $\Delta_{\sigma}$ defined in (\ref{A8}),  and
$E_{p,\sigma}(\rho)=e^{i\bar{p}\cdot\bar{x}}F_{n_{\sigma}}^{k}(u)$,
we arrive first at
\begin{eqnarray}\label{A25}
\lefteqn{\hspace{-1cm}\bigg[u^{2}\frac{\partial^{2}}{\partial
u^{2}}+u\frac{\partial}{\partial u}-\frac{u^{2}}{4}
}\nonumber\\
&&\hspace{-0.5cm}+\left(\frac{\hat{p}_{2}}{\alpha}+\frac{\sigma}{2}\right)u+
\frac{(p-\hat{p}_{2}^{2})}{\alpha^{2}}\bigg]F_{n_{\sigma}}^{k}(u)=0.
\end{eqnarray}
In the above equations, $\rho\equiv(t,u,y,z)$ where $u$ is a
dimensionless variable defined by $u\equiv
\frac{2}{\alpha^{2}}eB_{0}e^{-\alpha x}$. Moreover,
$\hat{p}_{2}=p_{2}+eB_{0}/\alpha$, and $(\bar{x},\bar{p})$ are given
as in the previous section [see below (\ref{A12})]. Comparing
(\ref{A25}) with the differential equation
\begin{eqnarray}\label{A26}
\lefteqn{\hspace{-1cm}\bigg[u^{2}\frac{\partial^{2}}{\partial
u^{2}}+u\frac{\partial}{\partial u}-\frac{u^{2}}{4}
}\nonumber\\
&&+\frac{(2n+k+1)}{2}u-\frac{k^{2}}{4}\bigg]\Phi_{n}^{k}(u)=0,
\end{eqnarray}
satisfied by
\begin{eqnarray}\label{A27}
\Phi_{n}^{k}(u)=\sqrt{\frac{n_{!}}{(n+k)!}}e^{-u/2}e^{k/2}{\cal{L}}_{n}^{k}(u),
\end{eqnarray}
where ${\cal{L}}_{n}^{k}(u)$ is the associated Laguerre polynomial,
the quantum numbers $k$ and $n_{\sigma}$ are given by
\begin{eqnarray}\label{A28}
k&=&\frac{2}{\alpha}\sqrt{\hat{p}_{2}^{2}-p},\qquad\mbox{and}\nonumber\\
n_{\sigma}&=&\frac{\sigma}{2}-\frac{1}{2}+\frac{\hat{p}_{2}-\sqrt{\hat{p}_{2}^{2}-p}}{\alpha}.
\end{eqnarray}
Note that the indices $k$ and $n_{\sigma}$ in
${\cal{L}}_{n_{\sigma}}^{k}$ are to be positive integers. This
implies the following quantization for $\hat{p}_{2}$ and $p$
\begin{eqnarray}\label{A29}
\hspace{-1cm}\lfloor\frac{\hat{p}_{2}}{\alpha}\rfloor\equiv s,\qquad
\mbox{and}\qquad p\equiv\alpha^{2}(s^{2}-r^{2}),
\end{eqnarray}
leading to $k=2r$, with $r>0$, and
\begin{eqnarray}\label{A29a}
n_{\sigma}=\frac{\sigma}{2}-\frac{1}{2}+s-r.
\end{eqnarray}
In (\ref{A29}), $\lfloor a\rfloor$ is the greatest integer less than
or equal to $a$. As it is shown in \cite{handrich2005, ghosh2009,
raya2010}, $s=\lfloor p_{2}/\alpha\rfloor$ is associated with
certain constant length $\ell_{0}\equiv \frac{e^{\alpha
x_{0}}}{\sqrt{eB_{0}}}$, through the relation
\begin{eqnarray}\label{A30}
s=\lfloor\frac{\hat{p}_{2}}{\alpha}\rfloor=\lfloor\frac{1}{(\alpha\ell_{0})^{2}}\rfloor,
\end{eqnarray}
where $x_{0}$ is a fixed length. At this stage, let us compare,
$n_{\sigma}$ from (\ref{A29a}) with $n_{\sigma}$ appearing in
(\ref{A18a}) for $eB_{0}>0$.\footnote{Comparing to (\ref{A18a}),
$n_{\sigma}$ from (\ref{A29a}), is in general given by
$n_{\sigma}=\frac{\sigma}{2}\mbox{sgn}(eB_{0})-\frac{1}{2}+s-r$.} As
it turns out, in the case of exponentially decaying magnetic field,
$s-r$ labels the energy levels. In analogy to the case of constant
magnetic field, to keep $n_{\sigma}$ from (\ref{A29a}) positive, the
electron spin orientation $\sigma$ in the lowest energy level,
characterized by $s-r=0$, is fixed to be positive ($\sigma=+1$). In
this case, we have $n\equiv n_{+}=s-r$. What concerns the quantum
numbers $r$ and $s$, for $r>0$, we get $s\geq r$, that guarantees
$n=s-r\geq 0$. In analogy to the solutions (\ref{A17}), for the
constant magnetic fields, we arrive therefore at
\begin{eqnarray}\label{A32}
\hspace{-0.8cm}F_{n}^{2r}(u)\equiv
F_{n_{+}}^{2r}(u)=N_{n,r}e^{-\frac{u}{2}}u^{r}{\cal{L}}_{n}^{2r}(u),
\end{eqnarray}
with $n=s-r=0,1,2,\cdots,$ for positive spin solutions and
\begin{eqnarray}\label{A33}
\hspace{-0.4cm}F_{n-1}^{2r}(u)\equiv F_{n_{-}}^{2r}(u)=
N_{n-1,r}e^{-\frac{u}{2}}u^{r}{\cal{L}}_{n-1}^{2r}(u),
\end{eqnarray}
with $n=s-r=1,2,3,\cdots$, for negative spin solutions. Using the
orthonormality relations of associated Legendre polynomials
${\cal{L}}_{n}^{2r}$, the normalization factor $N_{n,r}$ is given by
$$
N_{n,r}\equiv\sqrt{\frac{n!}{(n+ 2r)!}}.
$$
Using finally the projection matrices $\Delta_{\pm}=\frac{1}{2}(1\pm
i\gamma^{1}\gamma^{2})$, the solution for $\mathbb{E}_{n}$ can be
brought in the form similar to the Ritus eigenfunction (\ref{A13})
and (\ref{A14}) for constant magnetic field,
\begin{eqnarray}\label{A34}
\mathbb{E}_{p}(\rho)=e^{-i(p_{0}t-p_{2}y-p_{3}z)}P_{n}^{2r}(u),
\end{eqnarray}
with
\begin{eqnarray}\label{A35}
\lefteqn{P_{n}^{2r}(u)=}\nonumber\\
&&\hspace{-0.3cm}\frac{1}{2}\big\{[F_{n}^{2r}(u)+F_{n-1}^{2r}(u)]+
i\gamma^{1}\gamma^{2}[F_{n}^{2r}(u)-F_{n-1}^{2r}(u)]\big\}.\nonumber\\
\end{eqnarray}
Using the orthonormality conditions for the associated Laguerre
polynomials, ${\cal{L}}_{n}^{2r}(u)$, presented in App. \ref{appA},
it is straightforward to derive the closure and the orthonormality
relations for $\mathbb{E}_{p}$,
\begin{eqnarray}\label{A36}
\hspace{-0.5cm}\alpha\sum\limits_{r=1}^{\infty}\sum\limits_{s=r}^{\infty}
\int\frac{d^{2}{\mathbf{p}}_{\|}}{(2\pi)^{2}}
~\mathbb{E}_{p}(\rho)\bar{\mathbb{E}}_{p}(\rho')=\hat{\delta}(\rho-\rho'),
\end{eqnarray}
as well as
\begin{eqnarray}\label{A37}
\hspace{-0.5cm}\int{\cal{D}}\rho~
\mathbb{E}_{p}(\rho)\bar{\mathbb{E}}_{p'}(\rho)=
\alpha^{-1}\delta^{2}({\mathbf{p}}_{\|}-{\mathbf{p}}'_{\|})
\delta_{s,s'}\Pi_{r,r'}^{s}.
\end{eqnarray}
In the above relations, $\alpha\equiv\xi\sqrt{eB_{0}}$,
$\bar{\mathbb{E}}_{p}(\rho)\equiv
\gamma_{0}\mathbb{E}_{p}^{\dagger}(\rho)\gamma_{0}$,
${\mathbf{p}}_{\|}\equiv (p_{0},p_{3})$ and
$\hat{\delta}(\rho-\rho')\equiv
\delta(t-t')\delta(u-u')\delta(y-y')\delta(z-z')$,
${\cal{D}}\rho\equiv dt du dy dz$, and
\begin{eqnarray}\label{A38}
\lefteqn{\hspace{-0.3cm}\Pi_{r,r'}^{s}=\Delta_{+}\delta_{r,r'}\delta_{s,r}+(1-\delta_{s,r})\bigg[\delta_{r,r'}
}\nonumber\\
&&\hspace{-0.3cm}+\bigg\{\theta(r'-r)(-1)^{r'-r}
\sqrt{\frac{\Gamma(s+r)\Gamma(s-r)}{\Gamma(s+r')\Gamma(s-r')}}\nonumber\\
&&\hspace{-0.3cm}\times\left(\Delta_{+}\sqrt{\frac{(s^{2}-r^{2})}{(s^{2}-r'^{2})}}+\Delta_{-}\right)+r\to
r'\bigg\}\bigg],
\end{eqnarray}
with $\Delta_{\pm}$ given in (\ref{A8}). These relations are similar
to the results presented in \cite{raya2010}. Note that the term
proportional to $\delta_{r,s}$ isolates the contribution of the
lowest energy level, characterized by $s=r$. This term is also
proportional to $\Delta_{+}$, implying that the spin orientation in
the lowest energy level for $eB_{0}>0$ is positive. What concerns
the contribution of higher energy levels, we had to distinguish
between $r=r'$, $r>r'$ and $r<r'$ cases. This is because plugging
$\mathbb{E}_{p}$ from (\ref{A34}) and (\ref{A35}) in the l.h.s. of
(\ref{A37}), leads in general to the orthonormality relations
similar to (\ref{app1}) in App. \ref{appA}, where $r$ and $r'$ are
not necessarily equal. In App. \ref{appA}, the orthonormality
relation (\ref{app1}) is determined separately for $r=r'$, $r>r'$
and $r<r'$. Using these results, it is straightforward to derive the
remaining terms in (\ref{A38}), arising from higher energy levels.
\par
To determine the Ritus momentum $\tilde{p}$, we insert the above
solution (\ref{A34}) and (\ref{A35}) for $\mathbb{E}_{p}$ in
(\ref{A2}). After some straightforward algebraic manipulations,
where we use the recursion relations of the Laguerre polynomials
${\cal{L}}_{n}^{2r}(u)$, we arrive at
\begin{eqnarray}\label{A39}
\hspace{-0.7cm}(\gamma\cdot
\Pi)\mathbb{E}_{p}=\mathbb{E}_{p}[\gamma^{0}p_{0}+\gamma^{2}\alpha\sqrt{n(n+2r)}-\gamma^{3}p_{3}].
\end{eqnarray}
Here, $n=s-r$ and $\alpha=\xi\sqrt{eB_{0}}$. Comparing the r.h.s. of
(\ref{A39}) with (\ref{A2}), we get
$\tilde{p}=\left(p_{0},0,-\alpha\sqrt{n\left(n+2r\right)},p_{3}\right)$,
which can alternatively be given in terms of $(s,r)$ and $\xi$ as
\begin{eqnarray}\label{A40}
\tilde{p}_{\kappa}=\left(p_{0},
0,-\xi\sqrt{eB_{0}\kappa},p_{3}\right),
\end{eqnarray}
with $\kappa\equiv s^{2}-r^{2}$. Using (\ref{A40}), the energy
dispersion relation of a single electron in an exponentially
decaying magnetic field is given by
\begin{eqnarray}\label{A41}
{\cal{E}}_{\kappa}=\sqrt{\xi^{2} eB_{0}\kappa+p_{3}^{2}+m^{2}},
\end{eqnarray}
which is to be compared with the energy dispersion relation of an
electron in the presence of a constant magnetic field from
(\ref{A21}). Let us notice, that although $n=s-r$ labels the energy
levels, in contrast to the case of constant magnetic field,
$\kappa=s^{2}-r^{2}$ determines the ordering of energy levels.
Because of the above mentioned conditions for the integer quantum
numbers $r>0$, $s\geq r$ (or $n\geq 0$), some value of $\kappa$ are
not allowed. The simplest way to determine $\kappa$, is to write it
in terms of $n$ and $r$ as $\kappa=n(n+2r)$. Choosing
$r=1,2,\cdots$, then $\kappa$ turns out to be $\kappa=0$ for $n=0$
(or all values of $r=s$), $\kappa=3,5,7,\cdots, 1+2r$ for $n=1$,
$\kappa=8,12,16,\cdots, 2(2+2r)$ for $n=2$, etc. In Secs. \ref{sec3}
and \ref{sec4}, the summation over energy levels in an exponentially
decaying magnetic field will be replaced by
$\sum_{r=1}^{\infty}\sum_{s=r}^{\infty}g(r,s)$. Here, $g(r,s)$ is a
generic function, that depends on the quantum numbers $r$ and $s$.
According to the above descriptions, the lowest energy level is then
characterized by $\sum_{r=1}^{\infty}g(r,s=r)$, or equivalently by
$\sum_{r=1}^{\infty}g(r,n=0)$.
\par
Following the description presented at the end of the previous
section, the fermion propagator of an electron in an exponentially
decaying magnetic field is given by (see also \cite{raya2010})
\begin{eqnarray}\label{A42}
\lefteqn{G(\rho,\rho')\equiv
\langle\psi(\rho)\bar{\psi}(\rho')\rangle}\nonumber\\
&=&\alpha^{2}\sum\limits_{r=1}^{\infty}\sum\limits_{s=r}^{\infty}\int\frac{d^{2}{\mathbf{p}}_{\|}}{(2\pi)^{2}}~
\mathbb{E}_{p}(\rho)\frac{i}{(\gamma\cdot\tilde{p}_{\kappa}-m)}\bar{\mathbb{E}}_{p}(\rho'),\nonumber\\
\end{eqnarray}
where $\mathbb{E}_{p}(\rho)$ and $\tilde{p}_{\kappa}$ are given in
(\ref{A34})-(\ref{A35}) and (\ref{A40}), respectively. Using
(\ref{A2}) and the closure of $\mathbb{E}_{p}(\rho)$ from
(\ref{A36}), it is easy to verify
\begin{eqnarray}\label{T21a}
\left(\gamma\cdot \Pi-m\right)G(\rho-\rho')=\delta^{4}(\rho-\rho'),
\end{eqnarray}
where $\delta^{4}(\rho-\rho')\equiv
\alpha\delta(t-t')\delta(u-u')\delta(y-y')\delta(z-z')$. Here, as in
the case of constant magnetic fields, since we are working with
$eB_{0}>0$, the spin orientation in the lowest energy level is fixed
to be positive, and the negative spin electrons contribute only to
higher energy levels. These facts are explicitly implemented in the
choice $n=s-r=0,1,2,\cdots$ for positive spin solution
$F_{n}^{2r}(u)$ and $n=s-r=1,2,3,\cdots$ for negative spin solution
$F_{n-1}^{2r}(x)$ in (\ref{A32}) and (\ref{A33}),
respectively.\footnote{In other words, $F_{-1}^{2r}(u)$ is undefined
and is to be neglected, whenever it appears in a computation.}
\section{Fermion condensate $\langle\bar{\psi}\psi\rangle$ in external magnetic fields}\label{sec3}
\setcounter{equation}{0}\par\noindent In this section, we will
determine the fermion condensate $\langle\bar{\psi}\psi\rangle$ in
the presence of constant and exponentially decaying magnetic fields.
The fermion condensate is defined by \cite{schwinger1960}
\begin{eqnarray}\label{B1}
\langle\bar{\psi}\psi\rangle=-\lim\limits_{x\to
x'}\mbox{tr}[G_{F}(x-x')],
\end{eqnarray}
where $G_{F}(x-x')$ is the fermion propagator in the presence of an
external magnetic field, in general. In \cite{miransky1995},
Schwinger's proper-time formalism is used to determine
$\langle\bar{\psi}\psi\rangle$ in a constant magnetic field in the
LLL approximation. In this section, we will use the Ritus method,
and in particular, our results from the previous section to
determine $\langle\bar{\psi}\psi\rangle$ first in a constant and
then in an exponentially decaying magnetic field, in the LLL
approximation, at finite temperature and chemical potential.
\subsection{$\langle\bar{\psi}\psi\rangle$ in constant magnetic
fields}\label{sec3a}
\noindent To determine the fermion condensate in a constant magnetic
field, let us replace $G(\bar{x}-\bar{x}')$ from (\ref{A22}) in
(\ref{B1}). Performing the traces over $\gamma$-matrices, we arrive
at
\begin{eqnarray}\label{B2}
\langle\bar{\psi}\psi\rangle^{c}&=&-2im\sum\limits_{p=0}^{\infty}\int
\frac{dp_{0}dp_{2}dp_{3}}{(2\pi)^{3}}\frac{1}{(p_{0}^{2}-\omega_{p}^{2})}\nonumber\\
&&\times\{[f_{p}^{+}(x)]^{2}+[f_{p-1}^{+}(x)]^{2}\},
\end{eqnarray}
where according to (\ref{A20}), the $\omega_{p}$ for $eB_{0}>0$ is
defined by $\omega_{p}^{2}\equiv 2eB_{0}p+p_{3}^{2}+m^{2}$. In the
LLL approximation, we shall set $p=0$. Using the definition of
$f_{p}^{+}(x)$ from (\ref{A17}), and after performing the
integration over $p_{2}$,
\begin{eqnarray}\label{B3}
\int \frac{dp_{2}}{2\pi}[f_{0}^{+}(x)]^{2}=\frac{eB_{0}}{2\pi},
\end{eqnarray}
we arrive at
\begin{eqnarray}\label{B4}
\langle\bar{\psi}\psi\rangle^{c}_{\mbox{\tiny{LLL}}}=-\frac{imeB_{0}}{\pi}
\int\frac{dp_{0}dp_{3}}{(2\pi)^{2}}~\frac{1}{(p_{0}^{2}-\omega_{0}^{2})},
\end{eqnarray}
where $\omega_{0}^{2}\equiv p_{3}^{2}+m^{2}$. The dimensional
reduction into two longitudinal dimensions
${\mathbf{p}}_{\|}=(p_{0},p_{3})$, which occurs whenever the LLL
approximation is used \cite{miransky1995}, can be observed in
(\ref{B4}). To evaluate (\ref{B4}), we use
\begin{eqnarray}\label{B5}
\int\frac{d^{2}{\mathbf{p}}_{\|}}{(2\pi)^{2}}~\frac{1}{(\mathbf{p}_{\|}^{2}-m^{2})}\stackrel{\Lambda\gg~
m}{\longrightarrow}\frac{i}{4\pi}\ln\left(\frac{m^{2}}{\Lambda^{2}}\right),
\end{eqnarray}
which yields the fermion condensate in the presence of a constant
magnetic field in the lowest Landau level,
\begin{eqnarray}\label{B6}
\langle\bar{\psi}\psi\rangle_{\mbox{\tiny{LLL}}}^{c}=\frac{meB_{0}}{4\pi^{2}}
\ln\left(\frac{m^{2}}{\Lambda^{2}}\right)+{\cal{O}}(m).
\end{eqnarray}
Here, $\Lambda$ is an appropriate momentum cutoff. This result
coincides exactly with the result presented in \cite{miransky1995},
where the method of proper-time is used. Note that although no
integration over the coordinates $\bar{x}=(t,x,y,z)$ is performed
here, the condensate (\ref{B6}) is constant in $\bar{x}$. This is
mainly because of the special form of $f_{0}^{+}(x)$ from
(\ref{A17}) and arises from the integration over $p_{2}$ in
(\ref{B3}). In Sec. \ref{sec3b}, we will show that in the presence
of an exponentially decaying magnetic field, the condensate
$\langle\bar{\psi}\psi\rangle$ in the LLL depends nontrivially on
$x$. Let us also notice that in massless QED, the mechanism of
magnetic catalysis \cite{klimenko1992, miransky1995} is made
responsible for dynamically breaking of chiral symmetry and the
generation of a finite dynamical mass in the LLL. Moreover, as it is
shown in \cite{ferrer-anomalous}, apart from a dynamical mass
induced by the external magnetic field in the LLL, an anomalous
magnetic moment is also induced, so that the dynamical mass in the
LLL is to be replaced by the rest energy $E_{0}$, which is a
combination of the induced dynamical mass and the anomalous magnetic
moment in the LLL. In this paper, we do not consider the effects of
dynamically induced anomalous magnetic moment on the dynamically
generated fermion mass. The latter arises either as a solution of an
appropriate Schwinger-Dyson equation in a ladder approximation or by
solving a corresponding gap equation to an appropriate effective
potential in the mean field approximation. In what follows, we will
compute $\langle\bar{\psi}\psi\rangle_{\mbox{\tiny{LLL}}}^{c}$ from
(\ref{B6}) at finite temperature $T$ and finite density $\mu$. To do
this we use the standard imaginary time formalism, and replace
$p_{0}$ in (\ref{B4}) by $ip_{4}= i(\omega_{\ell}+i\mu)$, where
$\omega_{\ell}=\pi(2\ell+1)T$ is the Matsubara frequency labeled by
$\ell$. Thus the integration over $p_{0}$ is to be replaced by
\begin{eqnarray}\label{B7}
\int\frac{dp_{4}}{2\pi}\to
\frac{1}{\beta}\sum\limits_{\ell=-\infty}^{\infty},
\end{eqnarray}
where $\beta^{-1}\equiv T$.  The fermion condensate
$\langle\bar{\psi}\psi\rangle_{\mbox{\tiny{LLL}}}^{c}$ at finite $T$
and $\mu$ is therefore given by
\begin{eqnarray}\label{B8}
\langle\bar{\psi}\psi\rangle^{c}_{\mbox{\tiny{LLL}}}=-\frac{meB_{0}}{\pi\beta}
\sum\limits_{\ell=-\infty}^{\infty}\int\frac{dp_{3}}{2\pi}\frac{1}{(p_{4}^{2}+p_{3}^{2}+m^{2})}.\nonumber\\
\end{eqnarray}
To evaluate the integration over $p_{3}$ and the summation over
$\ell$, the relation
\begin{widetext}
\begin{eqnarray}\label{B9}
\lefteqn{\frac{1}{\beta}\sum\limits_{\ell=-\infty}^{\infty}\int\frac{d^{d}p}{(2\pi)^{d}}
\frac{({\mathbf{p}}^{2})^{a}p_{4}^{2t}}
{(p_{4}^{2}+\mathbf{p}^{2}+m^{2})^{\sigma}}=\frac{1}{2(4\pi)^{d/2}\Gamma(\sigma)\beta}
\frac{\Gamma\left(\frac{d}{2}+a\right)}{\Gamma\left(\frac{d}{2}\right)}\left(\frac{2\pi}{\beta}\right)^{-2\sigma+d+2(t+a)}
}\nonumber\\
&&\times\sum\limits_{k=0}^{\infty}\frac{(-1)^{k}}{k!}\Gamma\left(\sigma-a+k-\frac{d}{2}\right)\bigg[
\zeta\left(2\left(\sigma+k-t-a\right)-d;\frac{1}{2}-\frac{i\beta\mu}{2\pi}\right)
+(\mu\to -\mu)\bigg]\left(\frac{m\beta}{2\pi}\right)^{2k},
\end{eqnarray}
\end{widetext}
from \cite{sadooghi2009} will be used. This relation (\ref{B9}) is
originally derived in \cite{bedingham2000} for $\mu=0$. In
(\ref{B9}), $\zeta\left(s;a\right)$ is the Hurwitz zeta function
defined by $\zeta(s;a)=\sum_{k=0}^{\infty}(k+a)^{-s}$, where any
term with $k+a=0$ is excluded. In an evaluation up to
${\cal{O}}((m\beta)^{4})$, we get finally
\begin{eqnarray}\label{B10}
\lefteqn{\langle\bar{\psi}\psi\rangle^{c}_{\mbox{\tiny{LLL}}}=
-\frac{meB_{0}}{8\pi^{2}}\bigg\{2
\ln\left(\frac{m\beta}{2\pi}\right)-\gamma_{E}}\nonumber\\
&&-\bigg[\psi\left(\frac{1}{2}+\frac{i\beta\mu}{2\pi}\right)+
\psi\left(\frac{1}{2}-\frac{i\beta\mu}{2\pi}\right)\bigg]\nonumber\\
&&-\frac{(m\beta)^{2}}{8\pi^{2}}\bigg[\zeta\left(3;\frac{1}{2}+\frac{i\beta\mu}{2\pi}\right)+
\zeta\left(3;\frac{1}{2}-\frac{i\beta\mu}{2\pi}\right)\bigg]\bigg\}\nonumber\\
&&+{\cal{O}}((m\beta)^{4}),
\end{eqnarray}
where $\gamma_{E}\sim 0.557$ is the Euler-Mascheroni number and
$\psi(z)$ the polygamma function, defined by
$\psi(z)=\frac{d}{dz}\ln\Gamma(z)$. It arises by an appropriate
regularization of $\zeta(1;a)$ using
\begin{eqnarray}\label{B11}
\lim\limits_{\epsilon\to
0}\zeta(1+\epsilon;a)=\lim\limits_{\epsilon\to
0}\frac{1}{\epsilon}-\psi(a).
\end{eqnarray}
In (\ref{B10}), the divergent $1/\epsilon$ term is neglected, and
the identity $\psi(1/2)=-\gamma_{E}-2\ln 2$ is used.
\subsection{$\langle\bar{\psi}\psi\rangle$ in exponentially
decaying magnetic fields}\label{sec3b}
\noindent The chiral condensate in an exponentially decaying
magnetic field is given by (\ref{B1}), where $S_{F}(x-x')$ is to be
replaced by $G(\rho-\rho')$ from (\ref{A42}). Performing the trace
of Dirac matrices, we get first
\begin{eqnarray}\label{B12}
\langle\bar{\psi}\psi\rangle&=&-2i\xi^{2}meB_{0}\sum\limits_{r=1}^{\infty}\sum\limits_{s=r}^{\infty}
\int\frac{dp_{0}dp_{3}}{(2\pi)^{2}}~\frac{1}{(p_{0}^{2}-\omega_{\kappa}^{2})}\nonumber\\
&&\times \{[F_{n}^{2r}(u)]^{2}+[F_{n-1}^{2r}(u)]^{2}\},
\end{eqnarray}
where, according to (\ref{A40}), $\omega_{\kappa}^{2}\equiv
\xi^{2}eB_{0}\kappa+p_{3}^{2}+m^{2}$ and $\kappa=s^{2}-r^{2}$. The
expression on the r.h.s. of (\ref{B12}) includes the contributions
of all energy levels, denoted by $r$ and $s$, and the terms
proportional to $(F_{n}^{2r})^{2}$ and $(F_{n-1}^{2r})^{2}$
correspond to the contributions of electrons with positive and
negative spins, respectively. In the presence of strong magnetic
fields, assuming that the dynamics of the system is solely
determined by the LLL, the fermion condensate is given by
\begin{eqnarray}\label{B13}
\langle\bar{\psi}\psi\rangle_{\mbox{\tiny{LLL}}}&=&
-2i\xi^{2}meB_{0}e^{-u}[\cosh(u)-1]\nonumber\\
&&\times
\int\frac{dp_{0}dp_{3}}{(2\pi)^{2}}\frac{1}{(p_{0}^{2}-\omega_{0}^{2})},
\end{eqnarray}
in contrast to (\ref{B4}). The $u$-dependent factor behind the
integral arises by setting $n=0$ in (\ref{B12}) and using
${\cal{L}}_{0}^{2r}=1$ and
$$\sum\limits_{r=1}^{\infty}[F_{0}^{2r}(u)]^{2}=e^{-u}\sum\limits_{r=1}^{\infty}\frac{u^{2r}}{(2r)!}=e^{-u}\left(
\cosh{u}-1\right).$$ Note that in (\ref{B13}), the same dimensional
reduction to two dimensions, as in the case of constant magnetic
field, occurs. Evaluating the ${\mathbf{p}}_{\|}$-integration using
(\ref{B5}), we get
\begin{eqnarray}\label{B14}
\langle\bar{\psi}\psi\rangle_{\mbox{\tiny{LLL}}}&=&\frac{\xi^{2}meB_{0}}{2\pi}e^{-u}\left(\cosh
u-1\right)\ln\left(\frac{m^{2}}{\Lambda^{2}}\right)\nonumber\\
&&+{\cal{O}}(m).
\end{eqnarray} In contrast to (\ref{B6}), the
fermion condensate in an exponentially decaying magnetic field,
(\ref{B14}), depends on $u=\frac{2}{\xi^{2}}e^{-\xi\eta}$ with
$\eta\equiv x/\ell_{B}$, and $\ell_{B}=({eB_{0}})^{-1/2}$, the
magnetic length corresponding to $B_{0}$, the value of the magnetic
field at $x=0$. Following the same method to introduce finite
temperature and chemical potential as in the Sec. \ref{sec3a}, we
arrive at
\begin{eqnarray}\label{B15}
\lefteqn{\langle\bar{\psi}\psi\rangle_{\mbox{\tiny{LLL}}}=
-\frac{\xi^{2}meB_{0}}{4\pi}~e^{-u}(\cosh(u)-1)\bigg\{2
\ln\left(\frac{m\beta}{2\pi}\right)}\nonumber\\
&&-\gamma_{E}-\bigg[\psi\left(\frac{1}{2}+\frac{i\beta\mu}{2\pi}\right)+
\psi\left(\frac{1}{2}-\frac{i\beta\mu}{2\pi}\right)\bigg]\nonumber\\
&&-\frac{(m\beta)^{2}}{8\pi^{2}}\bigg[\zeta\left(3;\frac{1}{2}+\frac{i\beta\mu}{2\pi}\right)+
\zeta\left(3;\frac{1}{2}-\frac{i\beta\mu}{2\pi}\right)\bigg]\bigg\}\nonumber\\
&&+{\cal{O}}((m\beta)^{4}).
\end{eqnarray}
In Sec. \ref{sec5}, we will compare the chiral condensates
(\ref{B6}) and (\ref{B14}), as well as (\ref{B10}) and (\ref{B15}),
arising from constant and exponentially decaying magnetic fields,
respectively.
\section{Local electric current correlation function in external magnetic fields}\label{sec4}
\setcounter{equation}{0}
\par\noindent
In this section, we will determine the \textit{local} electric
current correlation function $\chi^{(i)}(x)$ in homogeneous and
inhomogeneous magnetic fields. It is defined generically by
\begin{eqnarray}\label{C1}
\chi^{(i)}(x)\equiv \lim\limits_{x'\to
x}\sum\limits_{n}\chi_{n}^{(i)}(x,x'),
\end{eqnarray}
with
\begin{eqnarray}\label{C2}
\lefteqn{\hspace{-0.7cm}\chi_{n}^{(i)}(x,x')}\nonumber\\
&&\hspace{-0.9cm}\equiv \int
{\cal{D}}{\hat{x}}{\cal{D}}\hat{x}'\mbox{tr}\big[\gamma^{i}G_{F}^{(n)}(\bar{x},\bar{x}')\gamma^{i}
G_{F}^{(n)}(\bar{x}',\bar{x})\big],
\end{eqnarray}
where $i=1,2,3$ are the spatial dimensions, and the sub- and
superscripts $n$ on $\chi_{n}^{(i)}$ and $G_{F}^{(n)}$ indicate the
contribution of each energy (Landau) level to the local electric
current correlation function $\chi^{(i)}(x)$ and the fermion
propagator $G_{F}$, respectively. Moreover, ${\cal{D}}\hat{x}\equiv
dtdydz$ and $\bar{x}=(t,x,y,z)$. Up to an integration over
$x$,\footnote{The $x$ direction is specified, because of the
specific Landau gauge, which fixes $A_{\mu}$ as is described in Sec.
\ref{sec2a}.} the above definition is consistent with the definition
of electric current susceptibility $\chi^{(i)}$ presented in
\cite{fukushima2009}. In what follows, we will replace
$G_{F}(\bar{x}-\bar{x}')$ in (\ref{C2}) by (\ref{A22}) and
(\ref{A42}) to determine the local electric current correlation
function in constant and exponentially decaying magnetic fields,
respectively.
\subsection{$\chi^{(i)}(x)$ in constant magnetic fields}\label{sec4a}
\noindent For a constant magnetic field, the contribution of
$\chi_{p}^{(i)}(x,x')$ corresponding to each Landau level, labeled
by $p$, is given by
\begin{eqnarray}\label{C3}
\chi^{(i)}_{p}(x,x')=\int
{\cal{D}}\hat{x}{\cal{D}}\hat{x}'\mbox{tr}\big[\gamma^{i}G_{p}(x,x')\gamma^{i}G_{p}(x',x)\big],\nonumber\\
\end{eqnarray}
where ${\cal{D}}\hat{x}=dtdydz$. Using (\ref{A22}) and plugging
\begin{eqnarray}\label{C4}
G_{p}(x,x')=\int{\cal{D}}\bar{p}~e^{i\bar{p}\cdot(\bar{x}-\bar{x}')}P_{p}(x)\frac{i}{\gamma\cdot
\tilde{p}_{p}-m}P_{p}(x'),\nonumber\\
\end{eqnarray}
with $P_{p}(x)$ from (\ref{A14}) and $\tilde{p}_{p}$ from
(\ref{A20}) in (\ref{C3}), we arrive after integrating over
coordinate and momentum variables, and taking the limit $x'\to x$,
first at
\begin{eqnarray}\label{C5}
\lefteqn{\chi_{p}^{(i)}(x)=\lim\limits_{x'\to
x}\chi_{p}^{(i)}(x,x')}\nonumber\\
&&=i\frac{L_{y}L_{z}}{T}\int\frac{{\cal{D}}\bar{p}}{(
p_{0}^{2}-\omega_{p}^{2})^{2}}\mbox{tr}[\gamma^{i}P_{p}(x)\left(\gamma\cdot
\tilde{p}_{p}+m\right)P_{p}(x) \nonumber\\
&&\ \ \times~
\gamma^{i}P_{p}(x)\left(\gamma\cdot\tilde{p}_{p}+m\right)P_{p}(x)].
\end{eqnarray}
Here, $L_{y}$ and $L_{z}$ are constant lengths of our probe in
second and third spatial directions. They arise from the integration
over $y$ and $z$, respectively. The temperature $T$ in (\ref{C5})
arises from the integration over the compactified imaginary time
$\tau\equiv it\in [0,\beta]$, where $\beta=T^{-1}$. Performing then
the trace over Dirac matrices, and summing over all Landau levels,
we get
\begin{eqnarray}\label{C6}
\hspace{-0.5cm}\chi^{(i)}(x)=2i\frac{L_{y}L_{z}}{T}\sum\limits_{p=0}^{\infty}\int\frac{dp_{0}dp_{2}dp_{3}}{(2\pi)^{3}}
\frac{{\cal{N}}_{p}^{(i)}}{(p_{0}^{2}-\omega_{p}^{2})^{2}},
\end{eqnarray}
where $\omega_{p}^{2}=\tilde{p}_{2}^{2}+p_{3}+m^{2}$ with
$\tilde{p}_{2}^{2}=2eB_{0}p$ for $eB_{0}>0$ [see (\ref{A20})], and
\begin{eqnarray}\label{C7}
{\cal{N}}_{p}^{(1)}&=&2(p_{0}^{2}-\omega_{p}^{2})[f_{p}^{+}(x)]^{2}[f_{p-1}^{+}(x)]^{2},\nonumber\\
{\cal{N}}_{p}^{(2)}&=&2(p_{0}^{2}-\omega_{p}^{2}+\tilde{p}_{2}^{2})
[f_{p}^{+}(x)]^{2}[f_{p-1}^{+}(x)]^{2},\nonumber\\
{\cal{N}}_{p}^{(3)}&=&
[p_{0}^{2}-\omega_{p}^{2}+2(p_{3}^{2}-\tilde{p}_{2}^{2})][f_{p}^{+}(x)]^{4}\nonumber\\
&&\hspace{-0.3cm}+
[p_{0}^{2}-\omega_{p}^{2}+2(p_{3}^{2}+\tilde{p}_{2}^{2})][f_{p-1}^{+}(x)]^{4}.
\end{eqnarray}
Note that in general the contribution in the transverse directions
$i=1,2$ are not equal. However, in the presence of strong magnetic
field, where the dynamics of the system is reduced to the lowest
Landau level, $p=0$, and we have therefore
\begin{eqnarray}\label{C8}
{\cal{N}}_{0}^{(1)}={\cal{N}}_{0}^{(2)}=0,
\end{eqnarray}
which lead to vanishing electric correlation functions in the
transverse directions, i.e. $\chi_{0}^{\perp}=0$. As concerns the
electric current correlation function in direction parallel to the
external magnetic field,  $\chi^{\|}_{0}$, plugging
\begin{eqnarray}\label{C9}
{\cal{N}}_{0}^{(3)}=(p_{0}^{2}-\omega_{0}^{2}+2p_{3}^{2})[f_{0}^{+}(x)]^{4},
\end{eqnarray}
from (\ref{C7}) in (\ref{C6}), using the definition of
$f_{0}^{+}(x)$ from (\ref{A17}), and integrating over $p_{2}$,
\begin{eqnarray}\label{C10}
\int\frac{dp_{2}}{2\pi}[f_{0}^{+}(x)]^{4}=\left(\frac{eB_{0}}{2\pi}\right)^{3/2},
\end{eqnarray}
we arrive at
\begin{eqnarray}\label{C11}
\chi_{0}^{\|}=\frac{2iL_{y}L_{z}}{T}\left(\frac{eB_{0}}{2\pi}\right)^{\frac{3}{2}}\int\frac{dp_{0}dp_{3}}{(2\pi)^{2}}
\frac{(p_{0}^{2}-\omega_{0}^{2}+2p_{3}^{2})}{(p_{0}^{2}-\omega_{0}^{2})^{2}}.\nonumber\\
\end{eqnarray}
Introducing the temperature as in the previous section, and using
(\ref{B9}) to perform the sum over $\ell$, labeling the Matsubara
frequencies in
\begin{eqnarray}\label{C12}
&&\hspace{-0.6cm}\frac{1}{\beta}\sum\limits_{\ell=-\infty}^{\infty}
\int\frac{dp_{3}}{2\pi}\bigg[\frac{1}{(p_{4}^{2}+p_{3}^{2}+m^{2})}-\frac{2p_{3}^{2}}{(p_{4}^{2}+p_{3}^{2}+m^{2})^{2}}
\bigg]\nonumber\\
&&=\frac{1}{4\pi},
\end{eqnarray}
we arrive finally at the components of the electric correlation
function in the transverse and longitudinal directions with respect
to a constant and strong magnetic field in the LLL approximation
\begin{eqnarray}\label{C13}
\chi_{0}^{\perp}&=&0,\nonumber\\
\chi_{0}^{\|}&=&\frac{L_{y}L_{z}}{2\pi
T}\left(\frac{eB_{0}}{2\pi}\right)^{3/2}.
\end{eqnarray}
Note that whereas $\chi_{0}^{\perp}=0$, the non-vanishing
$\chi_{0}^{\|}$ does not depend on $x$, although no integration over
$x$ is performed in the definition (\ref{C3}) of $\chi^{(i)}_{p}(x)$
from (\ref{C3}). To arrive at (\ref{C13}), the result from
(\ref{C12}) is used. In the chiral limit $m\to 0$, both integrals
appearing on the l.h.s. of (\ref{C12}) are infrared (IR) divergent,
and the (bare) fermion mass $m$, plays the role of an IR regulator.
Using the relation (\ref{B9}) to expand these integrals in a
high-temperature expansion, the IR divergences of the integrals
appearing in (\ref{C12}) cancel and we are left with an exact, $m$
independent solution for $\chi_{0}^{\|}$, which does not also depend
on the chemical potential $\mu$. Note that the above result
(\ref{C13}) suggests that for $eB_{0}\ll T^{2}$ and $L_{y}L_{z}\ll
\ell_{B}^{2}$ with $\ell_{B}=(eB_{0})^{-1/2},$ apart from
$\chi_{0}^{\perp}$, $\chi_{0}^{\|}$ also vanishes.
\par
The electric susceptibility of a massless magnetized QED at zero
temperature is recently studied in \cite{ferrer-para}. According to
the mechanism of magnetic catalysis \cite{klimenko1992,
miransky1995}, in massless QED, the exact chiral symmetry of the
original QED Lagrangian is dynamically broken by the external
magnetic field. In \cite{ferrer-para}, it is shown that the electric
susceptibility in the chirally broken phase at zero temperature is
independent of the applied magnetic field. It would be interesting
to extend the results presented in \cite{ferrer-para} to finite
temperature and study the temperature dependence of electric
susceptibility in chirally broken and symmetric phases of a
magnetized QED. This is indeed an open question and we hope to come
back to this point in the future.
\subsection{$\chi^{(i)}(u)$ in exponentially decaying magnetic fields}\label{sec4b}
\noindent In the case of exponentially decaying magnetic fields, the
dimensionless coordinate $u$ plays the same rule as $x$ in the
previous section. Thus, defining
\begin{eqnarray}\label{C14}
\chi_{\kappa}^{(i)}(u,u')=\int
{\cal{D}}\hat{x}{\cal{D}}\hat{x}'\mbox{tr}\big[\gamma^{i}G_{\kappa}(\rho,\rho')\gamma^{i}G_{\kappa}
(\rho',\rho)\big],\nonumber\\
\end{eqnarray}
in analogy to (\ref{C3}), using (\ref{A42}) and plugging
\begin{eqnarray}\label{C15}
\hspace{-1cm}G_{\kappa}(x,x')&=&\xi^{2}eB_{0}\int\frac{dp_{0}dp_{3}}{(2\pi)^{2}}~e^{i\bar{p}\cdot(\bar{x}-\bar{x}')}\nonumber\\
&&\hspace{-0.3cm}\times P_{n}^{2r}(u) \frac{i}{\gamma\cdot
\tilde{p}_{\kappa}-m}P_{n}^{2r}(u'),
\end{eqnarray}
with $P_{n}^{2r}(x)$ from (\ref{A35}) and $\tilde{p}_{\kappa}$ from
(\ref{A40}) in (\ref{C14}), we arrive after taking the limit $u'\to
u$ at
\begin{eqnarray}\label{C16}
\lefteqn{\hspace{-1cm}\chi_{\kappa}^{(i)}(u)=\lim\limits_{u'\to
u}\chi_{\kappa}^{(i)}(u,u')}\nonumber\\
&&\hspace{-1cm}=\frac{i\alpha^{3}L_{y}L_{z}}{T}\int\frac{dp_{0}dp_{3}}{(2\pi)^{2}}
\frac{1}{(p_{0}^{2}-\omega_{\kappa}^{2})^{2}}
\nonumber\\
&&\hspace{-0.5cm}\times
\mbox{tr}[\gamma^{i}P_{n}^{2r}(u)\left(\gamma\cdot
\tilde{p}_{\kappa}+m\right)P_{n}^{2r}(u') \nonumber\\
&&\hspace{-0.5cm}\times
\gamma^{i}P_{n}^{2r}(u')\left(\gamma\cdot\tilde{p}_{\kappa}+m\right)P_{n}^{2r}(u)],
\end{eqnarray}
where $\alpha=\xi\sqrt{eB_{0}}$, $n=s-r$, and, according to
(\ref{A40}), $\omega_{\kappa}^{2}=\tilde{p}_{2}^{2}+p_{3}^{2}+m^{2}$
with $\tilde{p}_{2}^{2}=\xi^{2}eB_{0}\kappa$ and
$\kappa=s^{2}-r^{2}$. After performing the trace over Dirac
matrices, the local electric current correlation function including
the contributions of all energy levels reads
\begin{eqnarray}\label{C17}
\chi^{(i)}(u)=
\frac{2i\alpha^{3}L_{y}L_{z}}{T}\sum\limits_{r=1}^{\infty}
\sum\limits_{s=r}^{\infty}\int\frac{dp_{0}dp_{3}}{(2\pi)^{2}}
\frac{{\cal{N}}^{(i)}_{r,s}}{(p_{0}^{2}-\omega_{\kappa}^{2})^{2}}.\nonumber\\
\end{eqnarray}
For different directions, $i=1,2,3$,  the nominator
${\cal{N}}_{r,s}^{(i)}$ is given by
\begin{eqnarray}\label{C18}
{\cal{N}}^{(1)}_{r,s}&=&2(p_{0}^{2}-\omega_{\kappa}^{2})[F_{n}^{2r}(u)]^{2}[F_{n-1}^{2r}(u)]^{2},\nonumber\\
{\cal{N}}^{(2)}_{r,s}&=&2(p_{0}^{2}-\omega_{\kappa}^{2}+2\tilde{p}_{2}^{2})[F_{n}^{2r}(u)]^{2}[F_{n-1}^{2r}(u)]^{2},\nonumber\\
{\cal{N}}^{(3)}_{r,s}&=&[p_{0}^{2}-\omega_{\kappa}^{2}+2(p_{3}^{2}-\tilde{p}_{2}^{2})]
[F_{n}^{2r}(u)]^{4}\nonumber\\
&&\hspace{-0.3cm}+[p_{0}^{2}-\omega_{\kappa}^{2}+2(p_{3}^{2}+\tilde{p}_{2}^{2})]
[F_{n-1}^{2r}(u)]^{4},
\end{eqnarray}
where $F_{n}^{2r}(u)$ is defined in (\ref{A32}). In the LLL, where,
according to our explanations in Sec. \ref{sec2b}, $s=r$ and
therefore $\kappa=0$, the only contribution to $\chi^{(i)}$ arises
from the positive spin particles. We get therefore
\begin{eqnarray}\label{C19}
{\cal{N}}^{(1)}_{r,s=r}={\cal{N}}^{(2)}_{r,s=r}=0,
\end{eqnarray}
leading to vanishing transverse components of the electric current
correlation function in the LLL approximation, i.e.
$\chi^{\perp}_{\mbox{\tiny{LLL}}}=0$. The same effect is also
observed in the case of constant magnetic fields in see Sec.
\ref{sec4a}. Moreover we get,
\begin{eqnarray}\label{C20}
{\cal{N}}^{(3)}_{r,s=r}=(p_{0}^{2}-\omega_{0}^{2}+2p_{3}^{2})[F_{0}^{2r}(u)]^{4}.
\end{eqnarray}
Plugging (\ref{C20}) in (\ref{C17}), we arrive first at
\begin{eqnarray}\label{C21}
\hspace{-1cm}\chi^{\|}_{\mbox{\tiny{LLL}}}(u)&=&
\frac{2i\alpha^{3}L_{y}L_{z}}{T}\sum\limits_{r=1}^{\infty}[F_{0}^{2r}(u)]^{4}\nonumber\\
&&\hspace{-1cm}\times \int\frac{dp_{0}dp_{3}}{(2\pi)^{2}}
\frac{(p_{0}^{2}-\omega_{0}^{2}+2p_{3}^{2})}{(p_{0}^{2}-\omega_{0}^{2})^{2}},
\end{eqnarray}
where $\|$ denotes the third (longitudinal) component of
$\chi^{(i)}$ in the LLL approximation. In (\ref{C21}), as in Sec.
\ref{sec3b}, the summation over $r$ can be performed using
\begin{eqnarray}\label{C22}
\sum\limits_{r=1}^{\infty}
\lefteqn{[F_{0}^{2r}(u)]^{4}=e^{-2u}\sum\limits_{r=1}^{\infty}\left(\frac{u^{2r}}{(2r)!}\right)^{2}
}\nonumber\\
&=& \frac{e^{-2u}}{2}\big[(I_{0}(2u)+J_{0}(2u))-2\big],
\end{eqnarray}
and ${\cal{L}}_{0}^{2r}=1$. Here, $I_{0}(z)$ and $J_{0}(z)$ are
zeroth order modified Bessel functions $I$ and $J$. Plugging
(\ref{C22}) in (\ref{C21}), we arrive at
\begin{eqnarray}\label{C23} \chi^{\|}_{\mbox{\tiny{LLL}}}(u)&=&
\frac{i\alpha^{3}L_{y}L_{z}}{T}e^{-2u}\big[(I_{0}(2u)+J_{0}(2u))-2\big]\nonumber\\
&&\times\int\frac{dp_{0}dp_{3}}{(2\pi)^{2}}
\frac{(p_{0}^{2}-\omega_{0}^{2}+2p_{3}^{2})}{(p_{0}^{2}-\omega_{0}^{2})^{2}}.
\end{eqnarray}
Introducing the temperature and finite density, using the method
introduced in the previous section, and using (\ref{C12}) to perform
the integration over $p_{3}$ and the sum over the Matsubara
frequencies, the transverse and longitudinal components of the local
electric current correlation function in the LLL approximation read
\begin{eqnarray}\label{C24}
\hspace{-0.5cm}\chi_{\mbox{\tiny{LLL}}}^{\perp}&=&0,\nonumber\\
\hspace{-0.5cm}\chi^{\|}_{\mbox{\tiny{LLL}}}(u)&=&\frac{\xi^{3}(eB_{0})^{3/2}~L_{y}L_{z}}{4\pi
T}\nonumber\\
&&\times e^{-2u}\big[I_{0}(2u)+J_{0}(2u)-2\big].
\end{eqnarray}
In the next section, we will compare local electric current
correlation functions (\ref{C13}) and (\ref{C24}) arising from
constant and exponentially decaying magnetic fields, and discuss the
remarkable property of (\ref{C24}) in the limit $\xi\to 0$.
\section{Discussions}\label{sec5}
\setcounter{equation}{0}
\par\noindent
In the previous sections, the chiral condensate
$\langle\bar{\psi}\psi\rangle$ and local electric current
correlation function $\chi^{(i)}$ are computed in the presence of
constant and exponentially decaying magnetic fields in the LLL
approximation. As it turns out the transverse ($i=1,2$) components
of $\chi^{(i)}$  vanish in this approximation. Moreover, whereas in
the presence of constant and strong magnetic fields
$\langle\bar{\psi}\psi\rangle_{\mbox{\tiny{LLL}}}^{c}$ from
(\ref{B6}) and $\chi^{\|}_{0}$ from (\ref{C13}) are constant (as a
function of coordinates), they depend, in the presence of
exponentially decaying magnetic fields, on a variable $u$ defined by
$u=\frac{2}{\xi^{2}}e^{-\xi \eta}$. Here, $\eta$ is proportional to
the coordinate $x$, and is given explicitly by $\eta=x/\ell_{B}$,
where the magnetic length $\ell_{B}=(eB_{0})^{-1/2}$. To give an
explicit example on $\eta$, let us take $eB_{0}=15 m_{\pi}^{2}\sim
0.3$ GeV$^{2}$, with the pion mass $m_{\pi}\sim 140$ MeV. This
corresponds to a magnetic field $B_{0}\sim 5\times 10^{19}$
Gau\ss,\footnote{We are working in the units where $eB=1$ GeV$^{2}$
corresponds to $B_{0}\sim 1.7\times 10^{20}$ Gau\ss~
\cite{sadooghi-sohrabi2009}.} which is the typical magnetic field
produced in the early stages of heavy ion collisions at LHC
\cite{mclerran2007} or exists in the interior of compact stars
\cite{incera2010}. In this case, the corresponding magnetic length
is given by $\ell_{B}\sim 0.4$ fm.\footnote{As it turns out, the
magnetic length $\ell_{B}$ for $eB_{0}=1$ GeV$^{2}$ is given by
$\ell_{B}\sim 0.63$ fm.} Taking $\eta=4$, for instance, would mean a
distance $x=4\ell_{B}\sim 1.6$ fm from the origin at $x=0$, etc.
Note that for the above mentioned LLL approximation to be reliable,
we have to consider only small damping parameter $\xi$ and consider
the probe in small $\eta$ from the origin $x=0$ with $B(x=0)=B_{0}$.
\begin{figure}[hbt]
\includegraphics[width=8cm,height=6cm]{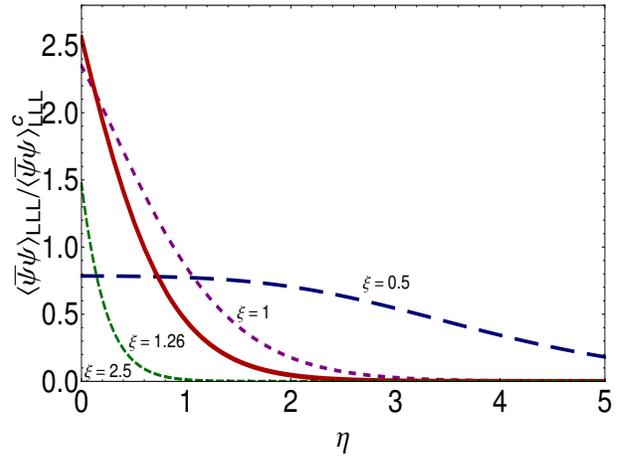}
\caption{The ratio ${\cal{C}}_{\xi}(\eta)={\langle
\bar{\psi}\psi\rangle_{\mbox{\tiny{LLL}}}}/{\langle
 \bar{\psi}\psi\rangle^{c}_{\mbox{\tiny{LLL}}}}$ is plotted as
a function of $\eta=x/\ell_{B}$ with $\ell_{B}=(eB_{0})^{-1/2}$ for
different damping parameters $\xi=0.5,1,1.26,2.5$. The condensates
$\langle\bar{\psi}\psi\rangle_{\mbox{\tiny{LLL}}}$ and
$\langle\bar{\psi}\psi\rangle_{\mbox{\tiny{LLL}}}^{c}$ are defined
in (\ref{B14}) and (\ref{B6}), respectively. For the damping factor
$\xi\sim 1.26$, the ratio ${\cal{C}}_{\xi}(\eta)$ is maximized at
$\eta=0$ (or equivalently $x=0$).}\label{fig1}
\end{figure}
\par
In this section, we will compare the values of
$\langle\bar{\psi}\psi\rangle_{\mbox{\tiny{LLL}}}^{c}$ from
(\ref{B6}) with $\langle\bar{\psi}\psi\rangle_{\mbox{\tiny{LLL}}}$
from (\ref{B14}), as well as $\chi^{(i)}_{0}$ from (\ref{C13}) with
$\chi^{(i)}_{\mbox{\tiny{LLL}}}$ from (\ref{C24}), in order to
explore the effect of inhomogeneity of the external magnetic field,
especially once quantum corrections are taken into account. To do
this, let us first define the ratio
\begin{eqnarray}\label{B16}
{\cal{C}}_{\xi}(\eta)\equiv
\frac{\langle\bar{\psi}\psi\rangle_{\mbox{\tiny{LLL}}}}
{\langle\bar{\psi}\psi\rangle_{\mbox{\tiny{LLL}}}^{c}}=2\pi\xi^{2}e^{-u}
\left(\cosh(u)-1\right),
\end{eqnarray}
with $u=\frac{2}{\xi^{2}}e^{-\xi \eta}$ and $\eta=x/\ell_{B}$.
\par
In Fig. \ref{fig1}, ${\cal{C}}_{\xi}(\eta)$ is plotted for different
damping parameters $\xi=0.5,1,1.26,2.5$ as a function of $\eta=
x/\ell_{B}$. Depending on the damping parameter $\xi$, the
condensate $\langle\bar{\psi}\psi\rangle_{\mbox{\tiny{LLL}}}$
arising from an exponentially decaying magnetic field is up to a
factor $2.5$ greater than the condensate
$\langle\bar{\psi}\psi\rangle_{\mbox{\tiny{LLL}}}^{c}$ in the
presence of a uniform magnetic field. For $\xi\sim 1.26$ the maximum
value of the condensate
$\langle\bar{\psi}\psi\rangle_{\mbox{\tiny{LLL}}}$ arises at
$\eta=0$ or equivalently at $x=0$. Let us notice that this
interesting observation is indeed in contradiction to our prior
expectation, according to which we expect ${\cal{C}}_{\xi}(\eta)=1$
at $\eta=0$ (or equivalently at $x=0$), because $B(x=0)=B_{0}$ is
constant. This indicates the singular nature of $\eta=0$ (or
equivalently $x=0$), once quantum fluctuations produce a
nonvanishing chiral condensate $\langle\bar{\psi}\psi\rangle$.
\begin{figure}[hbt]
\includegraphics[width=8cm,height=6cm]{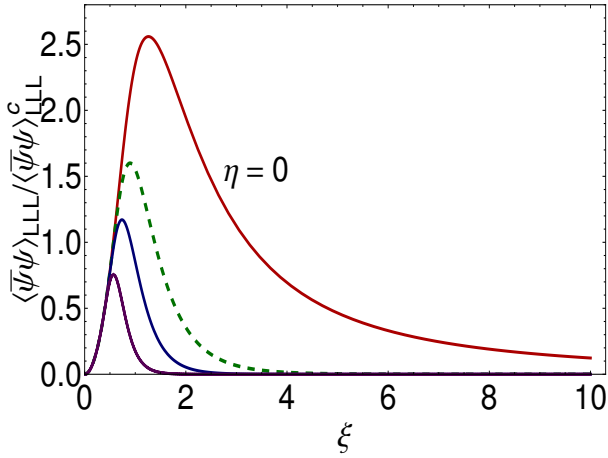}
\caption{The ratio ${\cal{C}}_{\xi}(\eta)={\langle
\bar{\psi}\psi\rangle_{\mbox{\tiny{LLL}}}}/{\langle
 \bar{\psi}\psi\rangle^{c}_{\mbox{\tiny{LLL}}}}$ from (\ref{B16}) is plotted as
a function of the damping factor $\xi$ for $\eta=0,0.5,1,2$ (or
equivalently for $x=0,\ell_{B}/2,\ell_{B},2\ell_{B}$ with
$\ell_{B}=(eB_{0})^{-1/2}$) from above to below. For $\eta=0$, the
${\cal{C}}_{\xi}(\eta)$ is maximized at $\xi\sim 1.26$. For
$\eta\neq 0$ (or equivalently $x\neq 0$), the maxima are shifted to
smaller values of $\xi$. For all values of $\eta$, the condensate
${\langle \bar{\psi}\psi\rangle_{\mbox{\tiny{LLL}}}}$, arising from
an exponentially decaying magnetic field, vanishes at
$\xi=0$.}\label{fig2}
\end{figure}
\begin{figure}[hbt]
\includegraphics[width=8cm,height=6cm]{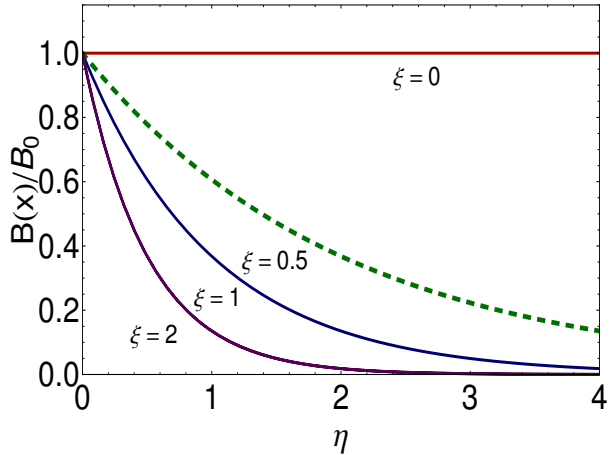}
\caption{The ratio $B(x)/B_{0}=e^{-\eta \xi}$ with $\eta=x/\ell_{B}$
and $\ell_{B}=(eB_{0})^{-1/2}$, is plotted as a function of the
damping factor $\eta$ for $\xi=0,0.5,1,2$ from above to below.
}\label{fig3}
\end{figure}
\par
The singular nature of ${\cal{C}}_{\xi}(\eta)$ for $\xi=0$ is
explored in Fig. \ref{fig2}, where the same ratio
${\cal{C}}_{\xi}(\eta)$ is plotted as a function of the damping
factor $\xi$ for $\eta=0, 0.5,1,2$ (or equivalently for
$x=0,\ell_{B}/2,\ell_{B},2\ell_{B}$). For $\eta=0$,
${\cal{C}}_{\xi}(0)$ is maximized at $\xi\sim 1.26$. For $\eta\neq
0$, however, the maxima of ${\cal{C}}_{\xi}(\eta)$ are shifted to
smaller values of $\xi$. At $\xi=0$, the condensate
$\langle\bar{\psi}\psi\rangle_{\mbox{\tiny{LLL}}}$, arising from an
exponentially decaying magnetic field, vanishes for all demonstrated
values of $\eta$. This is again in contradiction to our prior
expectation, according to which we expect ${\cal{C}}_{\xi}(\eta)=1$
for $\xi=0$. This is because for $\xi=0$ the exponentially decaying
magnetic field $B(x)=B_{0}e^{-\xi\eta}$ becomes constant (see Fig.
\ref{fig3}).
\begin{figure}[hbt]
\includegraphics[width=8cm,height=5.9cm]{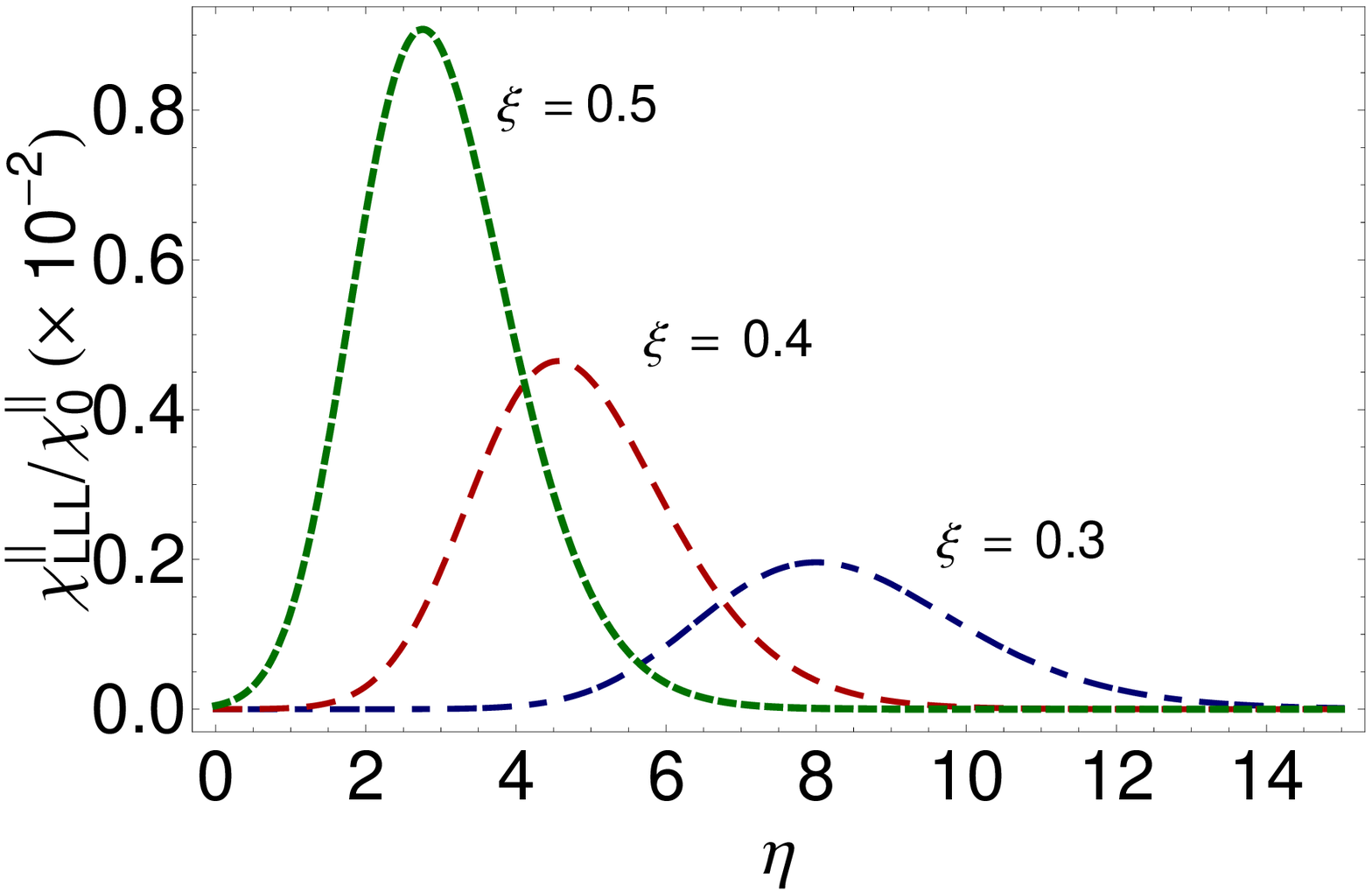}
\hspace{0.5cm}\includegraphics[width=8.4cm,height=5.9cm]{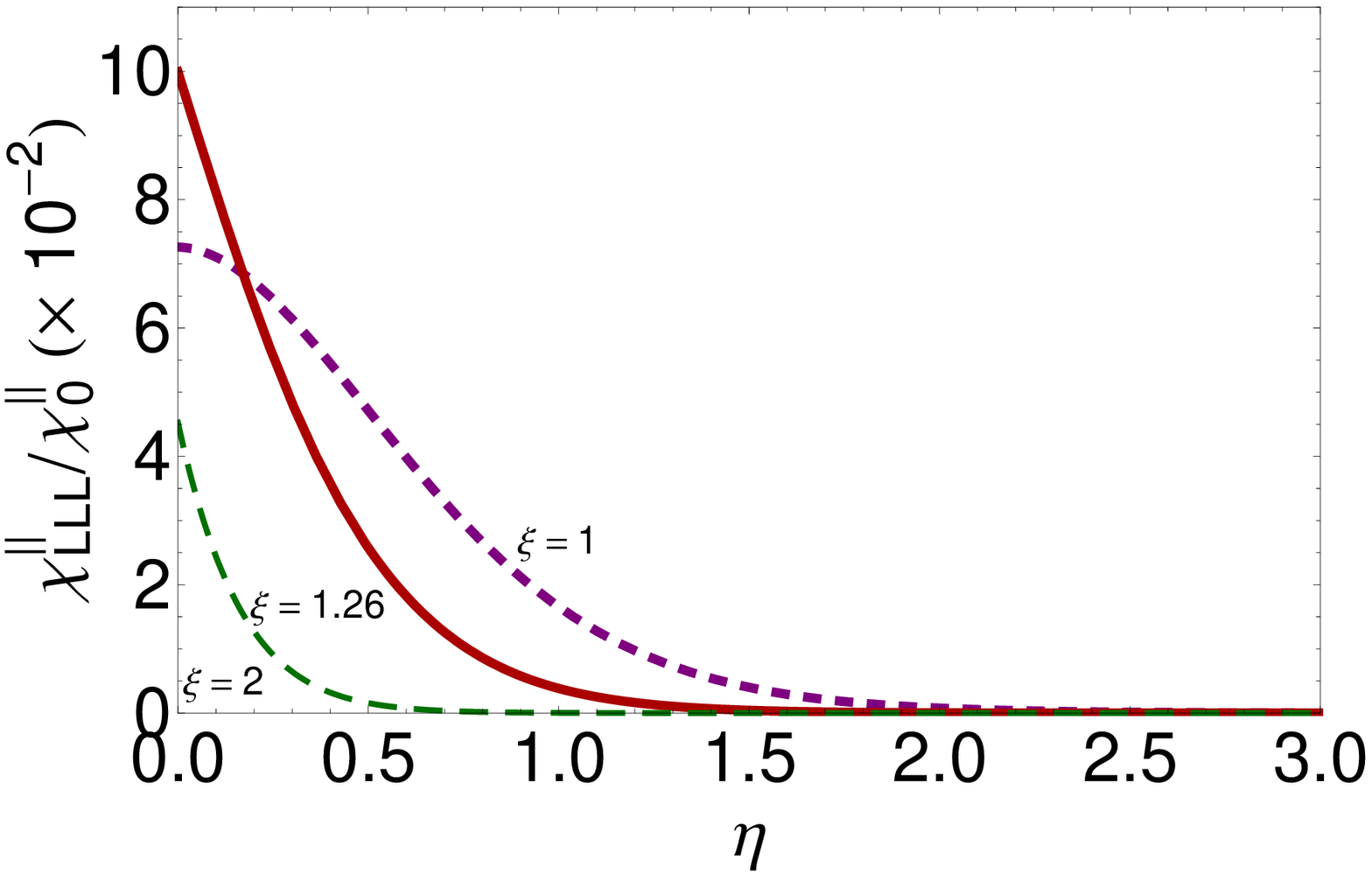}
\caption{The ratio
$\chi_{\xi}(\eta)={\chi_{\mbox{\tiny{LLL}}}^{\|}(u)}/{
 \chi_{0}^{\|}}$ from (\ref{C25}) is plotted as
a function of $\eta=x/\ell_{B}$ with $\ell_{B}=(eB_{0})^{-1/2}$ for
different damping parameters $\xi=0.1,0.2,\cdots,1,1.26,2$. For
$\xi\sim 1.26$, the ratio $\chi_{\xi}(\eta)$ is maximized at
$\eta=0$. }\label{fig4}
\end{figure}
\par
The same singular behavior at $\eta=0$ and $\xi=0$ is also observed
when we repeat the above analysis for the ratio
\begin{eqnarray}\label{C25}
\lefteqn{\chi_{\xi}(\eta)\equiv\frac{\chi_{\mbox{\tiny{LLL}}}^{\|}(u)}{\chi_{0}^{\|}}
}\nonumber\\
&&=\sqrt{2\pi^{3}}\xi^{3}e^{-2u}[I_{0}(2u)+J_{0}(2u)-2],
\end{eqnarray}
where $\chi_{0}^{\|}$ from (\ref{C13}) and
$\chi_{\mbox{\tiny{LLL}}}^{\|}(u)$ from (\ref{C24}) are local
electric current correlation functions in the presence of a uniform
and an exponentially decaying magnetic fields, respectively. Here,
as before, $u=\frac{2}{\xi^{2}}e^{-\xi\eta}$ and
$\eta={x}/{\ell_{B}}$, with $\ell_{B}=(eB_{0})^{-1/2}$.
\par
\begin{figure}[hbt]
\includegraphics[width=8cm,height=6cm]{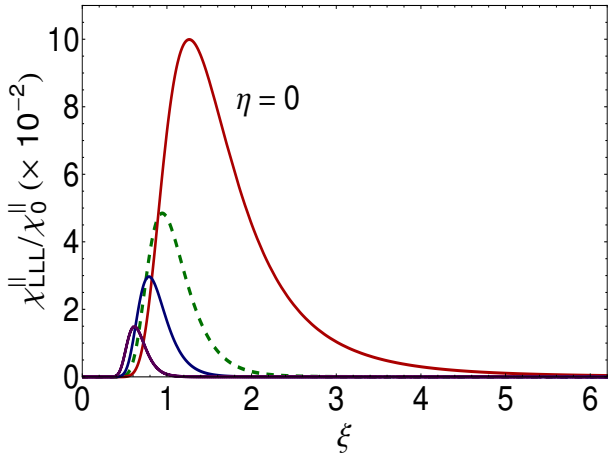}
\caption{ The ratio $\chi_{\xi}(\eta)$ from (\ref{C25}), is plotted
as a function of the damping factor $\xi$ for $\eta=0,0.5,1,2$  (or
equivalently for $x=0,\ell_{B}/2,\ell_{B},2\ell_{B}$ with
$\ell_{B}=(eB_{0})^{-1/2}$) from above to below. For $\eta=0$,
$\chi_{\xi}(\eta)$ is maximized at $\xi\sim 1.26$. For $\eta\neq 0$
(or equivalently $x\neq 0$), the maxima are shifted to smaller
values of $\xi$. For all values of $\eta$,
$\chi_{\mbox{\tiny{LLL}}}^{\|}(u)$ vanishes at $\xi=0$.
}\label{fig5}
\end{figure}

In Fig. \ref{fig4}, the ratio $\chi_{\xi}(\eta)$ is plotted for
$\xi=0.3,0.4,\cdots,2$ as a function of $\eta$. The fact that
$\chi_{\xi}(\eta)\neq 1$ for $\eta=0$ is again in contradiction to
our prior expectation: We know that the $x$-dependent magnetic field
$B(x)=B_{0}e^{-\eta \xi}$ becomes constant for $\eta=0$ (or
equivalently $x=0$), and therefore the prior expectation is that
$\chi_{\xi}(\eta)= 1$ as well as ${\cal{C}}_{\xi}(\eta)=1$ for
$\eta=0$. The above observation demonstrates again the singular
nature of the limit $x\to 0$, once the quantum effects are to be
considered. Another remarkable point here is that, according to Fig.
\ref{fig2}, for $\eta=0$ (or equivalently $x=0$) and in the interval
$0.5\lesssim \xi\lesssim 4$, the ratio ${\cal{C}}_{\xi}(\eta)\geq
1$, whereas according to Fig. \ref{fig4} (see also Fig. \ref{fig5}),
$\chi_{\xi}(\eta)\leq 1$ for $\eta=0$ and in the same interval
$0.5\lesssim \xi\lesssim 4$. In other words, for $\eta=0$ and in the
interval $0.5\lesssim \xi\lesssim 4$,\footnote{As before mentioned,
the LLL approximation is only reliable for small values of $\xi$ and
$\eta$.} the value of the electric current correlation function
$\chi^{\|}_{0}$ arising from a constant magnetic field is always
larger than $\chi^{\|}_{\mbox{\tiny{LLL}}}(u)$ arising from an
exponentially decaying magnetic field, whereas in the same interval
of $\xi$,  $\langle\bar{\psi}\psi\rangle_{\mbox{\tiny{LLL}}}^{c}$
arising from a constant magnetic field is always smaller than
$\langle\bar{\psi}\psi\rangle_{\mbox{\tiny{LLL}}}$ arising from an
exponentially decaying magnetic field.
\par
In Fig. \ref{fig5}, the same ratio $\chi_{\xi}(\eta)$ from
(\ref{C25}) is plotted as a function of the damping factor $\xi$ for
fixed $\eta=0,0.5,1,2$ (or equivalently for
$x=0,\ell_{B}/2,\ell_{B},2\ell_{B}$ with $\ell_{B}=(eB_{0})^{-1/2}$)
from above to below. As it turns out the value of
${\chi_{\mbox{\tiny{LLL}}}^{\|}}(u)$ for $\xi\to 0$ vanishes, in
contrast to our expectation.
\par
\begin{figure}[hbt]
\includegraphics[width=8.5cm,height=6.5cm]{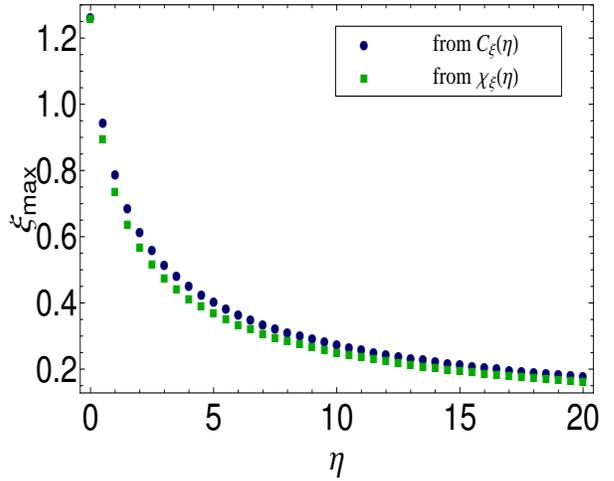}
\caption{The values of $\xi$ that maximize the ratios
${\cal{C}}_{\xi}(\eta)$ from (\ref{B16}) [blue circles] and
$\chi_{\xi}(\eta)$ from (\ref{C25}) [green rectangles] are plotted
as a function of $\eta$. They almost overlap. }\label{fig6}
\end{figure}

The second observation from Fig. \ref{fig5} is that the maxima of
$\chi_{\xi}(\eta)$ are shifted to smaller values of $\xi$ for
increasing $\eta$. In Fig. \ref{fig6}, the values of $\xi$ that
maximize the ratios defined in (\ref{B16}) [blue circles] and
(\ref{C25}) [green rectangles] are plotted as a function of $\eta$.
They are almost the same. This indicates a certain relation between
the formation of chiral condensates and the value of electric
current correlation function in a system including relativistic
fermions.
\par
\begin{figure*}[hbt]
\includegraphics[width=8cm,height=6cm]{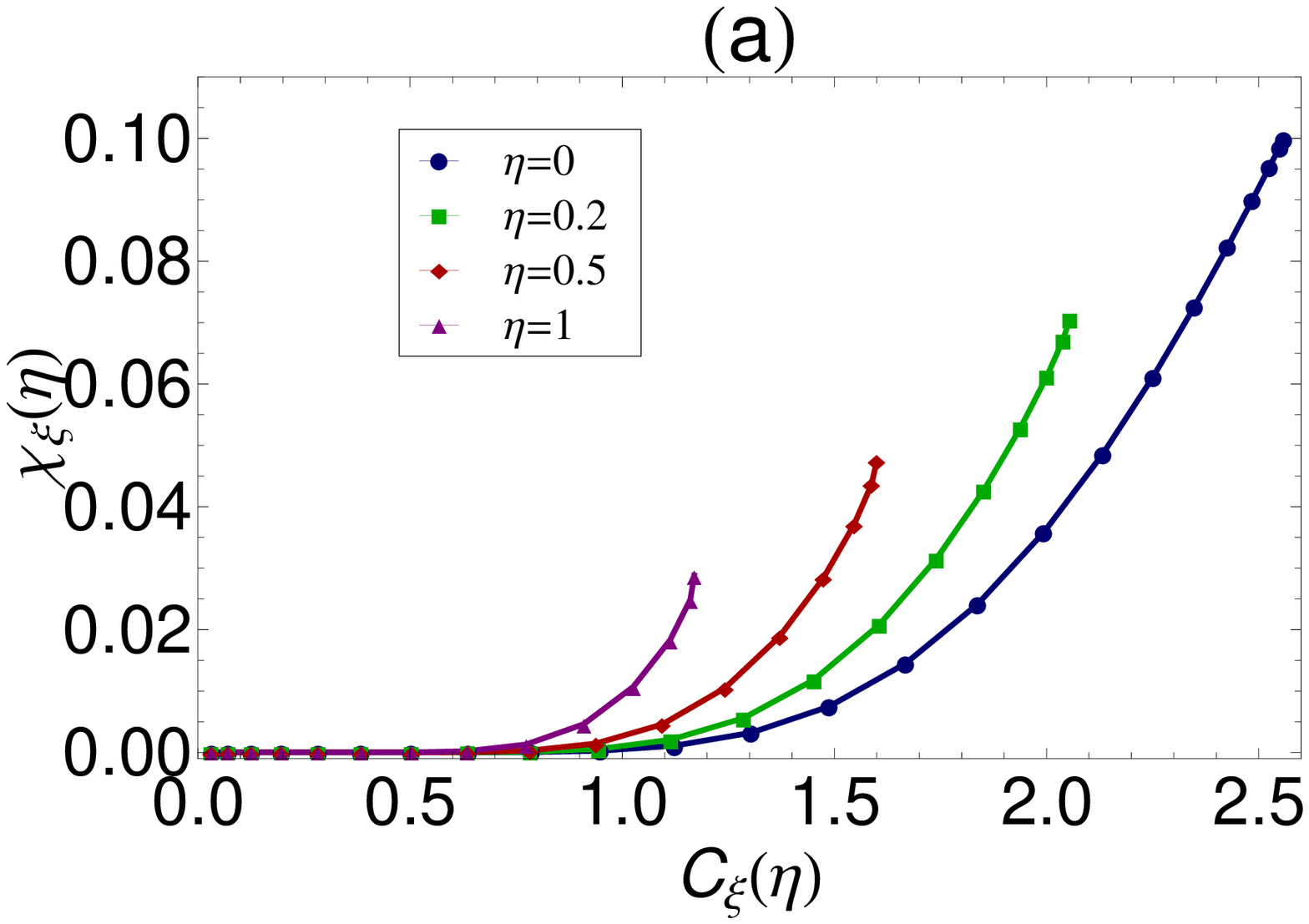}
\includegraphics[width=8cm,height=6cm]{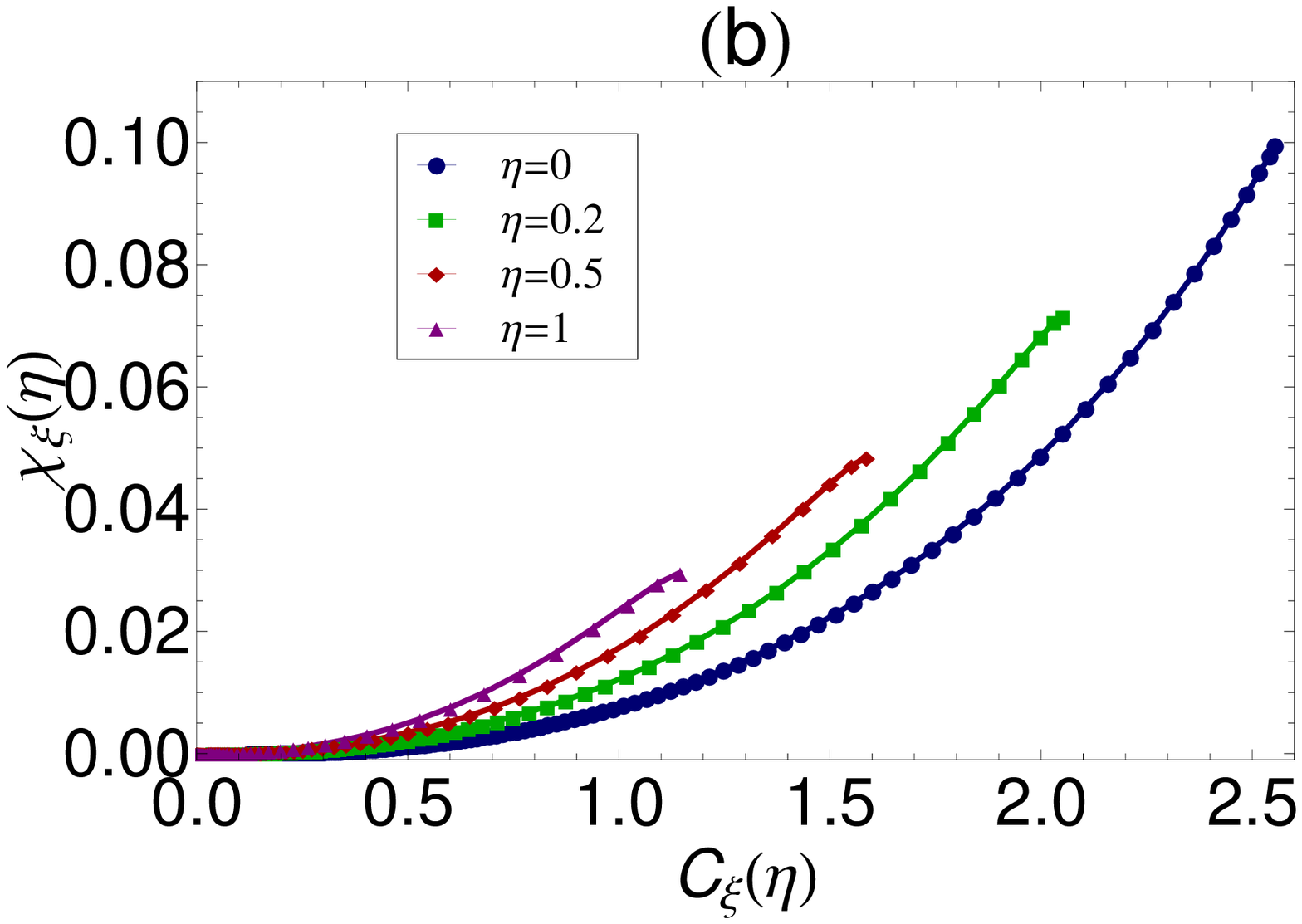}
\includegraphics[width=8cm,height=6cm]{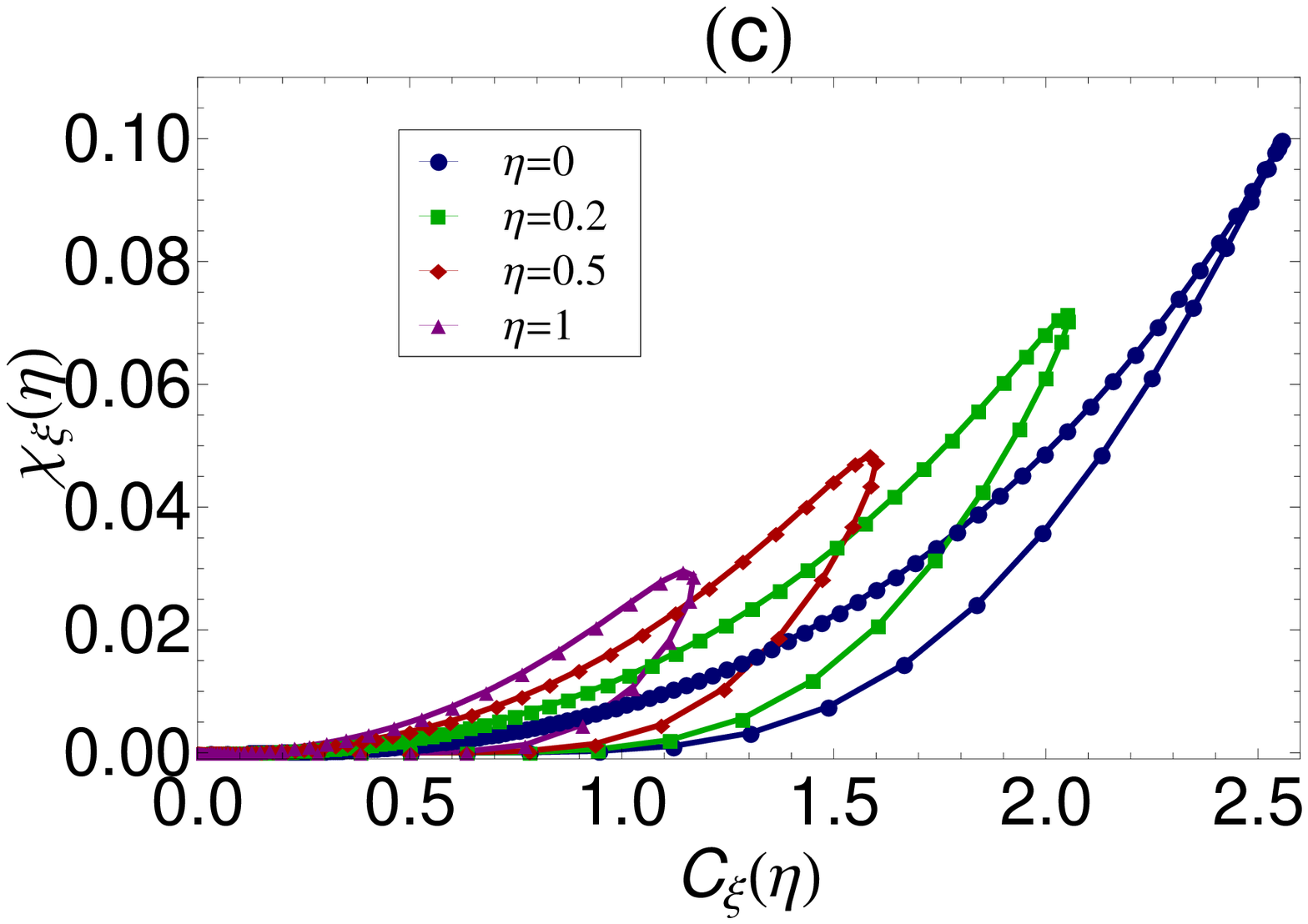}
\caption{The ratio $\chi_{\xi}(\eta)$ from (\ref{C25}) is plotted as
a function of the ratio ${\cal{C}}_{\xi}(\eta)$ from (\ref{B16}) for
fixed $\eta=0,0.2,0.5,1$ (or equivalently, $x=0, \ell_{B}/5,
\ell_{B}/2,\ell_{B}$) and various damping parameters
$\xi=0.1,\cdots, 10$. Each dot in the above curves indicates a value
of $\xi=0.1,\cdots,10$ with $\Delta\xi=0.05$. For $\xi$ smaller than
a certain $\xi_{\star}$, $\chi_{\xi}(\eta)$ increases with
increasing  ${\cal{C}}_{\xi}(\eta)$ (panel a). The value of
$\xi_{\star}$ is different for different $\eta$ (see the main text).
For $\xi>\xi_{\star}$, however, $\chi_{\xi}(\eta)$ decreases with
decreasing ${\cal{C}}_{\xi}(\eta)$ (panel b). The two curves in
panels a and b are reassembled in panel c. In the LLL approximation,
for small values of $\xi$ and $\eta$, only the behavior demonstrated
in panel a is relevant.}\label{fig7}
\end{figure*}

This relation is demonstrated in Figs. \ref{fig7}a-\ref{fig7}c,
where $\chi_{\xi}(\eta)$ from (\ref{C25}) is plotted as a function
of ${\cal{C}}_{\xi}(\eta)$ from (\ref{B16}) for fixed
$\eta=0,0.2,0.5,1$ (or equivalently, $x=0, \ell_{B}/5,
\ell_{B}/2,\ell_{B}$) and various damping parameters
$\xi=0.1,\cdots, 10$. Each dot in Figs. \ref{fig7}a-\ref{fig7}c
indicates a value of $\xi=0.1,\cdots,10$ with $\Delta\xi=0.05$. As
it turns out, $\chi_{\xi}(\eta)$ increases with increasing
${\cal{C}}_{\xi}(\eta)$ for a certain numbers of $\xi$ (see Fig.
\ref{fig7}a). The maximum value of $\xi$ for which this is true, is
different for different $\eta$. Let us denote this specific $\xi$
with $\xi_{\star}$. For $\eta=0, 0.2,0.5,1$, $\xi_{\star}$'s are
$\xi_{\star}\sim 1.3, 1.1, 0.95, 0.3$, respectively. For
$\xi>\xi_{\star}$, however, $\chi_{\xi}(\eta)$ decreases with
decreasing ${\cal{C}}_{\xi}(\eta)$ (see Fig. \ref{fig7}b). In Fig.
\ref{fig7}c, the two curves in Fig. \ref{fig7}a and \ref{fig7}b are
reassembled and demonstrate a hysteresis-like curve with increasing
$\xi$. However, since the LLL approximation is only valid for small
$\xi$ and $\eta$, we shall limit us to the values of $\xi\leq
\xi_{\star}$ for which $\chi_{\xi}(\eta)$ increases with increasing
${\cal{C}}_{\xi}(\eta)$ (Fig. \ref{fig7}a).
\section{Concluding Remarks}\label{sec6}
\par\noindent
In the present paper, the effect of exponentially decaying magnetic
fields on the dynamics of Dirac fermions is explored in detail.
Using the Ritus eigenfunction method, we have first determined the
energy spectrum of fermions in this non-uniform magnetic field and
compared it with the energy spectrum of relativistic fermions in a
constant magnetic field. We have then computed the chiral condensate
$\langle\bar{\psi}\psi\rangle$ and local electric current
correlation function $\chi^{(i)}$ in the presence of strong uniform
and non-uniform magnetic fields in the LLL approximation. In
non-uniform magnetic fields, $\langle\bar{\psi}\psi\rangle$ and
$\chi^{(i)}$ depend on a dimensionless variable
$u=\frac{2}{\xi^{2}}e^{-\xi x/\ell_{B}}$, which is a nontrivial
function of the magnetic field's damping parameter $\xi$, the
coordinate $x$ and the magnetic length $\ell_{B}=(eB_{0})^{-1/2}$.
In the LLL approximation, the transverse components of the electric
susceptibility, $\chi^{(i)}_{\mbox{\tiny{LLL}}}, i=1,2$, vanish and
its longitudinal component $\chi^{\|}_{\mbox{\tiny{LLL}}}$ depends
also on the variable $u$, only in the presence of exponentially
decaying magnetic fields. We have shown that the limits $\xi\to 0$
as well as $x\to 0$ are singular. In these limits, the $x$-dependent
magnetic field becomes constant, and therefore
$\langle\bar{\psi}\psi\rangle_{\mbox{\tiny{LLL}}}$ as well as
$\chi^{\|}_{\mbox{\tiny{LLL}}}$ arising from the $x$-dependent
magnetic field are expected to have the same value as in a constant
magnetic field. But, as it turns out this is not the case. This
remarkable behavior of
$\langle\bar{\psi}\psi\rangle_{\mbox{\tiny{LLL}}}$ as well as
$\chi^{\|}_{\mbox{\tiny{LLL}}}$ at $\xi\to 0$ and/or $x\to 0$ is
discussed in detail in Sec. \ref{sec5}. Let us notice that,
mathematically, there is a difference between taking the limit of a
quantity to zero and setting it exactly equal to zero. This
difference is indeed responsible for the singular behavior of the
limits $\xi\to 0$ and/or $x\to 0$. When we are looking for the
solution of the differential equation (\ref{A25}) for non-uniform
magnetic fields, for instance, the parameter $\xi$ appearing in $u=
\frac{2}{\xi^{2}}e^{-\xi\sqrt{eB_{0}}x}$ can be very small ($\xi\to
0$), but it cannot be set exactly equal to zero. When $\xi=0$, the
magnetic field becomes constant, and the differential equation
(\ref{A25}), leading to the solution (II.31), including associated
Laguerre polynomials, has to be replaced by the differential
equation (\ref{A16}) for constant magnetic fields, leading to
Hermite polynomials. There is no way to reproduce the Hermite
solution of the latter case from the associated Laguerre solution of
the former case, by taking $\xi\to 0$. This is why that although at
the classical level the uniform magnetic field is reproduced in
these limits from the non-uniform magnetic field, neither the energy
spectrum nor the quantum corrections to
$\langle\bar{\psi}\psi\rangle_{\mbox{\tiny{LLL}}}$ as well as
$\chi^{\|}_{\mbox{\tiny{LLL}}}$ of the constant magnetic fields can
be reproduced from their values in the non-uniform magnetic field by
taking the limit $\xi\to 0$.
\par
We have also shown that for small values of $x/\ell_{B}$ and $\xi$,
where we believe that the LLL approximation is reliable, the ratio
$\chi_{\xi}(\eta)$ from (\ref{C25}) increases with increasing ratio
${\cal{C}}_{\xi}(\eta)$ from (\ref{B16}) up to a maximum value of
$\xi$, denoted by $\xi_{\star}$. As it turns out, the value of
$\xi_{\star}$ depends on $\eta$ (see Sec. \ref{sec5}).
\par
Let us notice that electric susceptibility (electric current
correlation function) of a three-flavor color superconductor is
recently computed in a strong homogeneous magnetic field in
\cite{ferrer2012}, and the magnetoelectric effect in a strongly
magnetized quark matter is studied in detail. The present work is a
second nontrivial example on the effect of external magnetic fields,
in general, and spatially decaying magnetic fields, in particular,
on electric current correlation function of a system containing
relativistic Dirac fermions. Apart from applications in condensed
matter physics, the results of this paper may also be relevant in
the context of heavy ion collisions. Here, recent experimental
activities at RHIC indicate the production of intense magnetic
fields in the early stage of non-central heavy ion collisions
\cite{mclerran2007}. Depending on the initial conditions, e.g. the
energy of colliding nucleons and the corresponding impact
parameters, the spatially varying magnetic fields are estimated to
be in the order of $B\sim 10^{18}-10^{19}$ Gau\ss,~decaying very
fast, within few femtometers to $B\sim 10^{13}-10^{14}$ Gau\ss,~and
vanishing during the hadronization process \cite{mclerran2007}.

\section{Acknowledgments}
\noindent The authors acknowledge valuable discussions with Sh.
Fayazbakhsh, S. A. Jafari and A. Vaezi.
\begin{appendix}
\section{Orthonormality relations of associated Laguerre
polynomials}\label{appA} \setcounter{equation}{0}
\par\noindent
In this appendix, we present a number of useful orthonormality
relations of associated Laguerre polynomials. To do this, let us
define the integral
\begin{eqnarray}\label{app1}
I_{n,n'}^{r,r'}=\int_{0}^{\infty}du~
e^{-u}u^{r+r'}{\cal{L}}_{n}^{2r}(u){\cal{L}}_{n'}^{2r'}(u).
\end{eqnarray}
For $r=r'$, the standard orthonormality relation of associated
Laguerre polynomials is given by
\begin{eqnarray}\label{app2}
I_{n,n'}^{r,r}=\frac{(n+2r)!}{n!}\delta_{n,n'}.
\end{eqnarray}
But, in the present paper, we have to consider the cases, where in
general $r\neq r'$. To determine $I_{n,n'}^{r,r'}$, we have to
distinguish several cases, which are summarized in Tables
\ref{table2} and \ref{table3}.
\begin{widetext}
\begin{table*}[htb]
\begin{tabular}{ccccccccccc}
          \hline\hline
\multicolumn{7}{c}{$n>n'$ and $r'>r$}&&\multicolumn{3}{c}{$n>n'$ and $r'<r$}\\
$~n-n'>r'-r~$&~~&$~n-n'=r'-r~$&~~&$~n-n'<r'-r\leq n~$&~~&$~n<r'-r~$&~$\hspace{1.5cm}$~&$~r-r'\leq n'~$&~~&$~r-r'>n'~$\\
\hline
$I_{1}$&~~&$I_{2}$&~~&$I_{3}$&~~&$I_{4}$&~$\hspace{1.5cm}$~&$I_{5}$&~~&$I_{6}$\\
\hline\hline
\end{tabular}
\caption{The integrals $I_{i}, i=1,\cdots,6$ are presented in
(\ref{app3}) and (\ref{app4}).}\label{table2}
\end{table*}

\begin{table*}[htb]
\begin{tabular}{ccccccccccc}
          \hline\hline
\multicolumn{7}{c}{$n<n'$ and $r'<r$}&&\multicolumn{3}{c}{$n<n'$ and $r'>r$}\\
$~n'-n>r-r'~$&~~&$~n'-n=r-r'~$&~~&$~n'-n<r-r'\leq
n'~$&~~&$~n'<r-r'~$&~$\hspace{1.5cm}$~&$~r'-r< n~$&~~&$~r'-r\geq n~$\\
\hline
$I_{7}$&~~&$I_{8}$&~~&$I_{9}$&~~&$I_{10}$&~$\hspace{1.5cm}$~&$I_{11}$&~~&$I_{12}$\\
\hline\hline
\end{tabular}
\caption{The integrals $I_{i}, i=7,\cdots,12$ are presented in
(\ref{app5}) and (\ref{app6}).}\label{table3}
\end{table*}
The integrals $I_{i}, i=1,\cdots, 4$ for $n>n'$ and $r'>r$ are given
by
\begin{eqnarray}\label{app3}
I_{1}&=&0,\nonumber\\
I_{2}&=&\frac{(-1)^{r'-r}~(r'+r+n')!}{n' !}, \nonumber\\
I_{3}&=&\sum\limits_{\ell=0}^{(r'-r)-(n-n')}\frac{(-1)^{n-n'+\ell}~(r'-r)(r'+r+\ell-n')!~(r'-r+\ell-1)!}
{\ell!~(n'-\ell)!~(n-n'+\ell)!~(r'-r-n+n'-\ell)!},\nonumber\\
I_{4}&=&\sum\limits_{\ell=0}^{n'}\frac{(-1)^{n-n'+\ell}(r'-r)(r'+r+\ell-n')!(r'-r+\ell-1)!}
{\ell!~(n'-\ell)!~(n-n'+\ell)!~(r'-r-n+n'-\ell)!}.
\end{eqnarray}
For $n>n'$ and $r'<r$, we have
\begin{eqnarray}\label{app4}
I_{5}&=&\sum\limits_{\ell=0}^{r-r'}\frac{(-1)^{\ell}(r-r')(r'+r+n'-\ell)!(r-r'+n-n'+\ell-1)!}{\ell!~(n'-\ell)!(n-n'+\ell)!(r-r'-\ell)!},\nonumber\\
I_{6}&=&\sum\limits_{\ell=0}^{n'}\frac{(-1)^{\ell}(r-r')(r'+r+n'-\ell)!(r-r'+n-n'+\ell-1)!}{\ell!~(n'-\ell)!(n-n'+\ell)!(r-r'-\ell)!}.
\end{eqnarray}
The integrals $I_{i}, i=7,\cdots,10$ for $n<n'$ and $r'<r$ are given
by
\begin{eqnarray}\label{app5}
I_{7}&=&0,\nonumber\\
I_{8}&=&\frac{(-1)^{r-r'}~(r'+r+n)!}{n !}, \nonumber\\
I_{9}&=&\sum\limits_{\ell=0}^{(r-r')-(n'-n)}\frac{(-1)^{n'+\ell-n}~(r-r')(r+r'-n+\ell)!
~(r-r'+\ell-1)!}{\ell!~(n-\ell)!~(n'-n+\ell)!~(r-r'-n'+n-\ell)!},\nonumber\\
I_{10}&=&\sum\limits_{\ell=0}^{n}\frac{(-1)^{n'-n+\ell}(r-r')(r+r'+\ell-n)!(r-r'+\ell-1)!}
{\ell!~(n-\ell)!~(n'-n+\ell)!~(r-r'-n'+n-\ell)!}.
\end{eqnarray}
For $n<n'$ and $r'>r$, we get
\begin{eqnarray}\label{app6}
I_{11}&=&\sum\limits_{\ell=0}^{r'-r}\frac{(-1)^{\ell}(r'-r)(r'+r+n-\ell)!(r'-r+n'-n+\ell-1)!}{\ell!~(n-\ell)!(n'-n+\ell)!(r'-r-\ell)!},\nonumber\\
I_{12}&=&\sum\limits_{\ell=0}^{n}\frac{(-1)^{\ell}(r'-r)(r'+r+n-\ell)!(r'-r+n'-n+\ell-1)!}
{\ell!~(n-\ell)!(n'-n+\ell)!(r'-r-\ell)!}.
\end{eqnarray}
\end{widetext}
\end{appendix}

\end{document}